\documentclass{svmult}

\usepackage{makeidx}     
\usepackage{graphicx}    
\usepackage{multicol}    

\makeindex             

\begin{document}

\title*{Generalized Technical Analysis.\\ Effects of transaction volume and risk} 
\toctitle{Generalized Technical Analysis. Effects of transaction volume and risk}

\titlerunning{Generalized Technical Analysis. Effects of transaction volume and risk}

\author{Marcel Ausloos\inst{1}  \and Kristinka Ivanova\inst{2}}

\authorrunning{ M. Ausloos and K. Ivanova}

\institute{ GRASP, B5, University of Li$\grave e$ge, B-4000 Li$\grave e$ge,
Euroland 
\and Pennsylvania State University, University Park PA 16802, USA }

\maketitle

\begin{abstract}

We  generalize the momentum indicator idea  taking into account the volume
of transactions as a multiplicative factor. We compare  returns obtained following  strategies based
on the classical  or the generalized technical analysis, taking into account a sort of  risk  investor criterion.
\end{abstract}


\section{Introduction}

First we recall classical technical analysis methods of stock
evolution. We recall the notion of moving averages and momentum indicators. We
present a generalization of momentum indicators based on classical physics
principles, taking into account not only the price of a stock but also the volume
of transactions. Next we compare the returns obtained following  strategies based
on the classical technical analysis and the generalized technical analysis. The
cases of four stocks quoted on NASDAQ, four stocks traded on NYSE, three major
financial indices and the price of Gold will serve as illustrations. We consider
the volume of transactions and the daily closing price of these stocks and
indices for the period Jan. 01, 1997 to Dec. 31, 2001. Daily closing price signals 
$y(t)$ are plotted in Fig.1(a-d) for stocks quoted on NASDAQ, i.e. 
CSCO, SUNW, AMAT, and MSFT; in Fig.1(e-h) for stocks traded on the  NYSE, i.e. 
GE, AOL, PFE, and GFI; in Fig.1(i-k) are three financial indices: (i) NASDAQ, 
(j) S\&P500), (k) DJIA; the price of  Gold is in Fig.1 (l).

\begin{figure} \centering
\includegraphics[width=.30\textwidth]{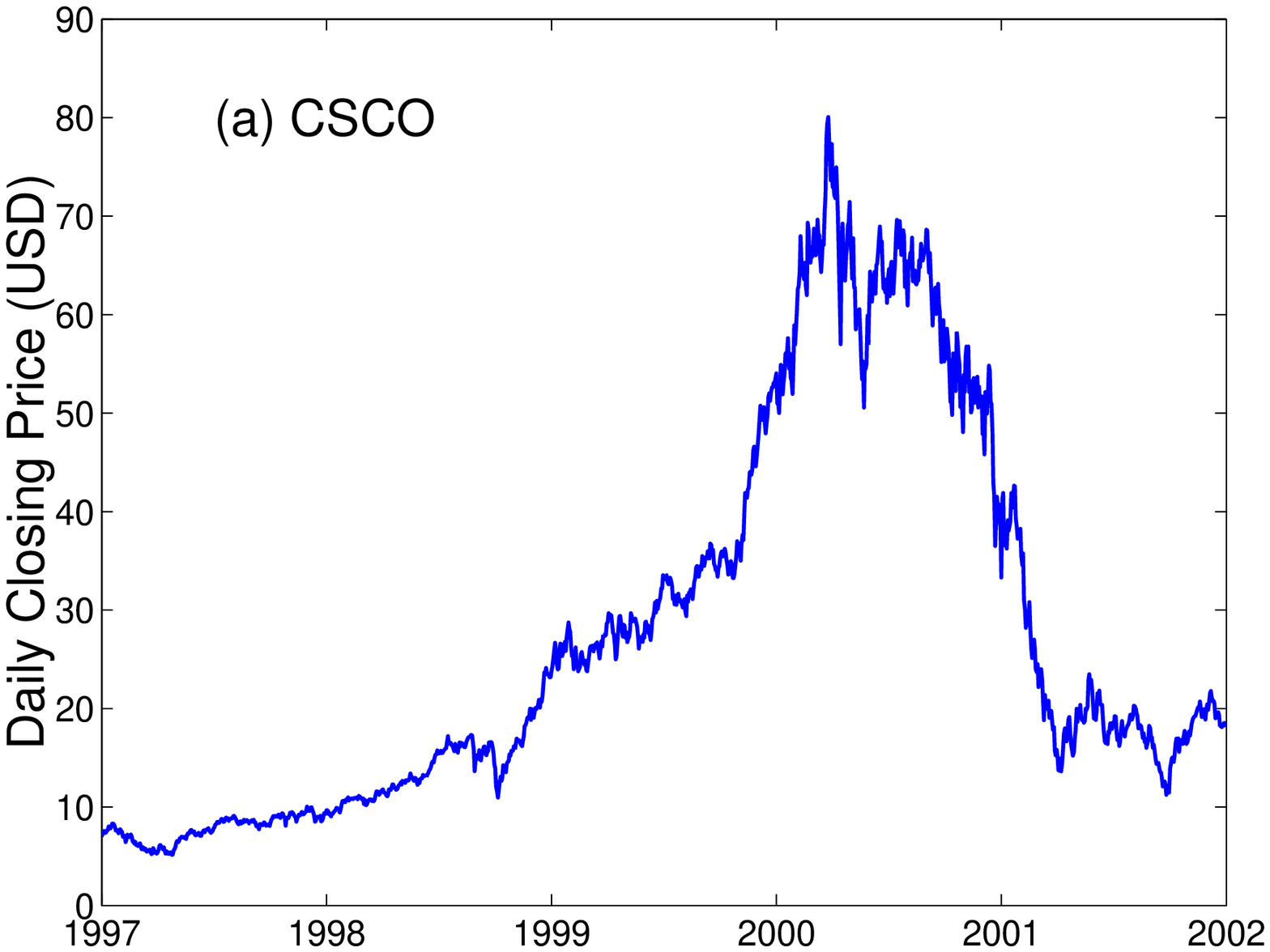} \hfill
\includegraphics[width=.30\textwidth]{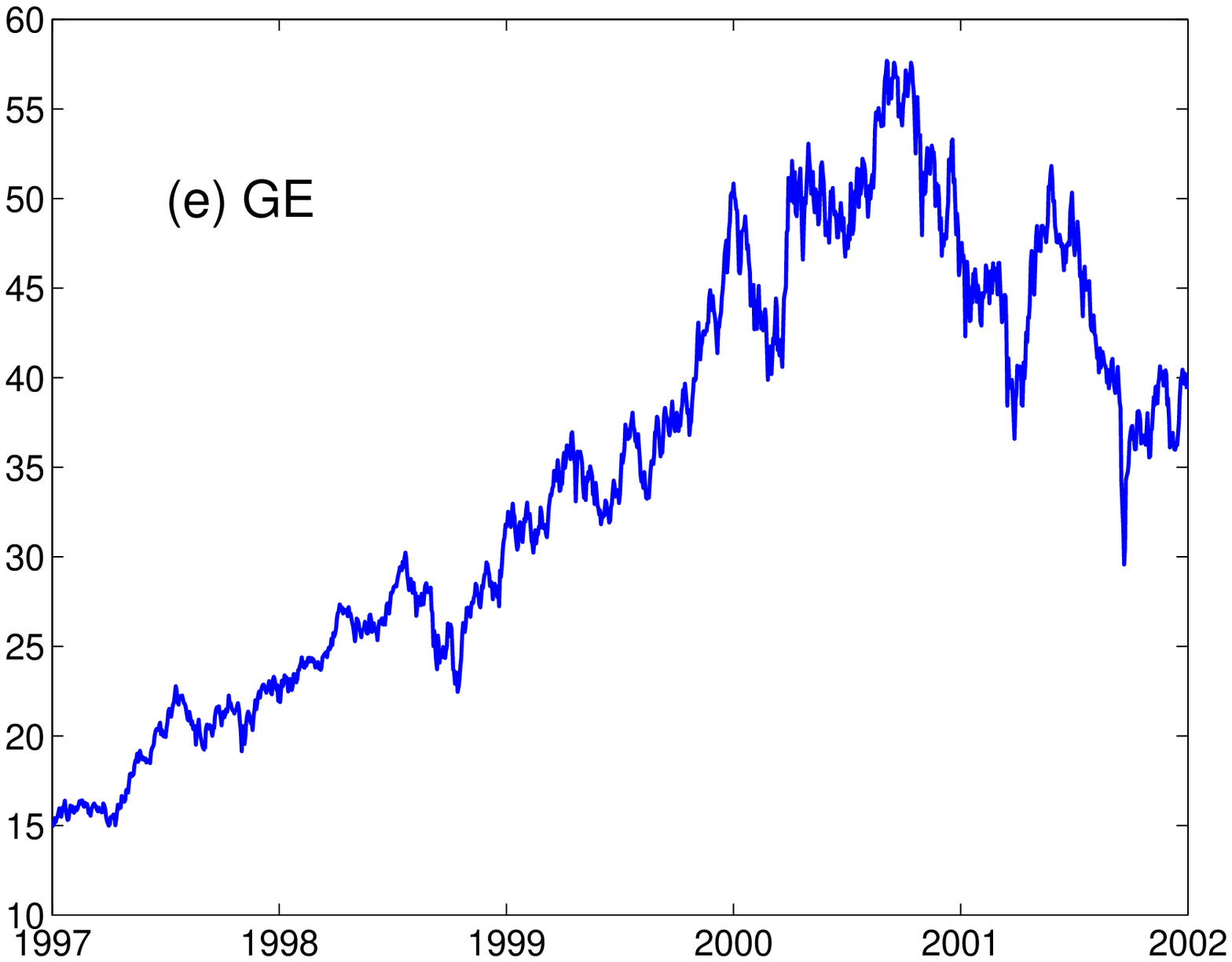} \hfill
\includegraphics[width=.30\textwidth]{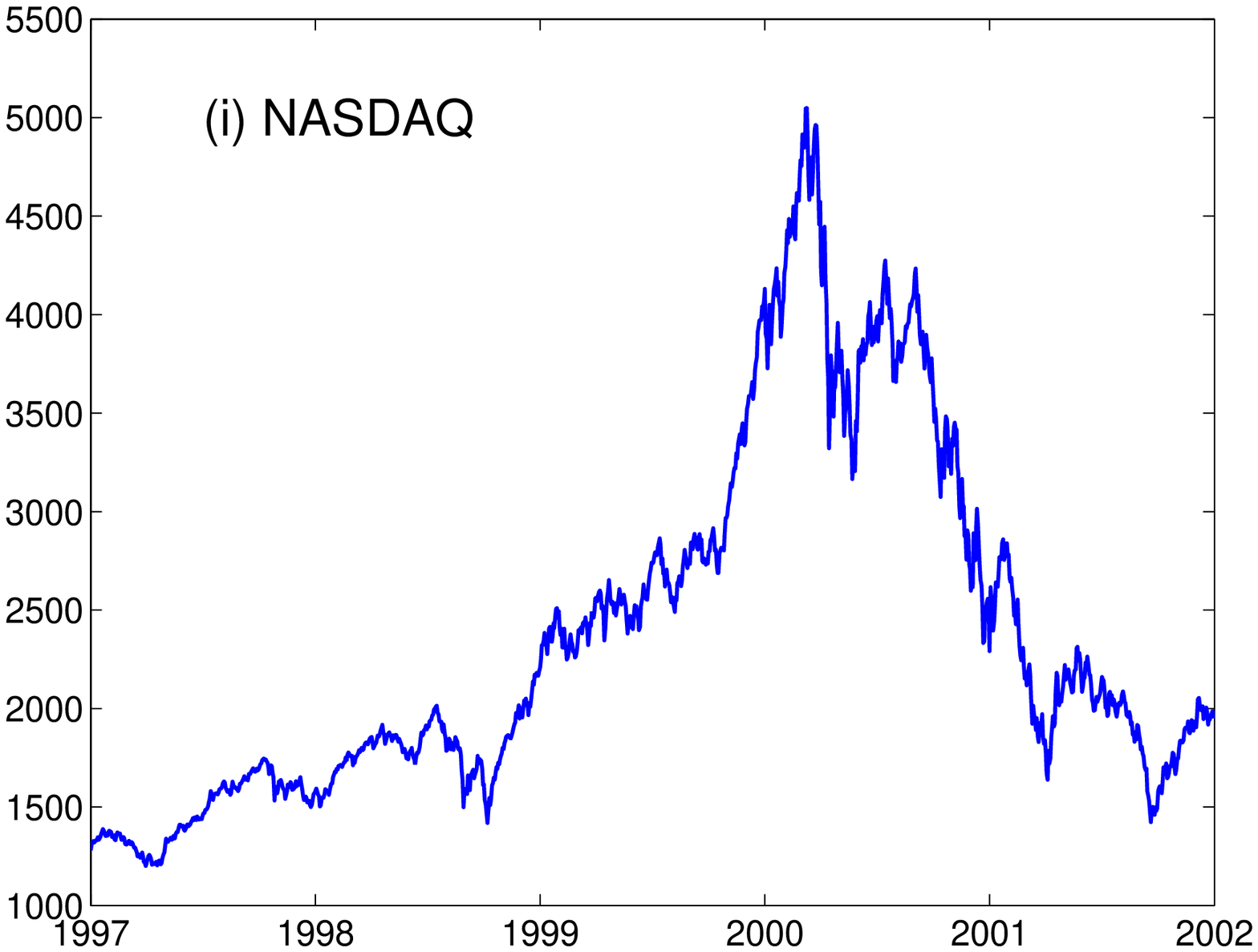} \vfill
\includegraphics[width=.30\textwidth]{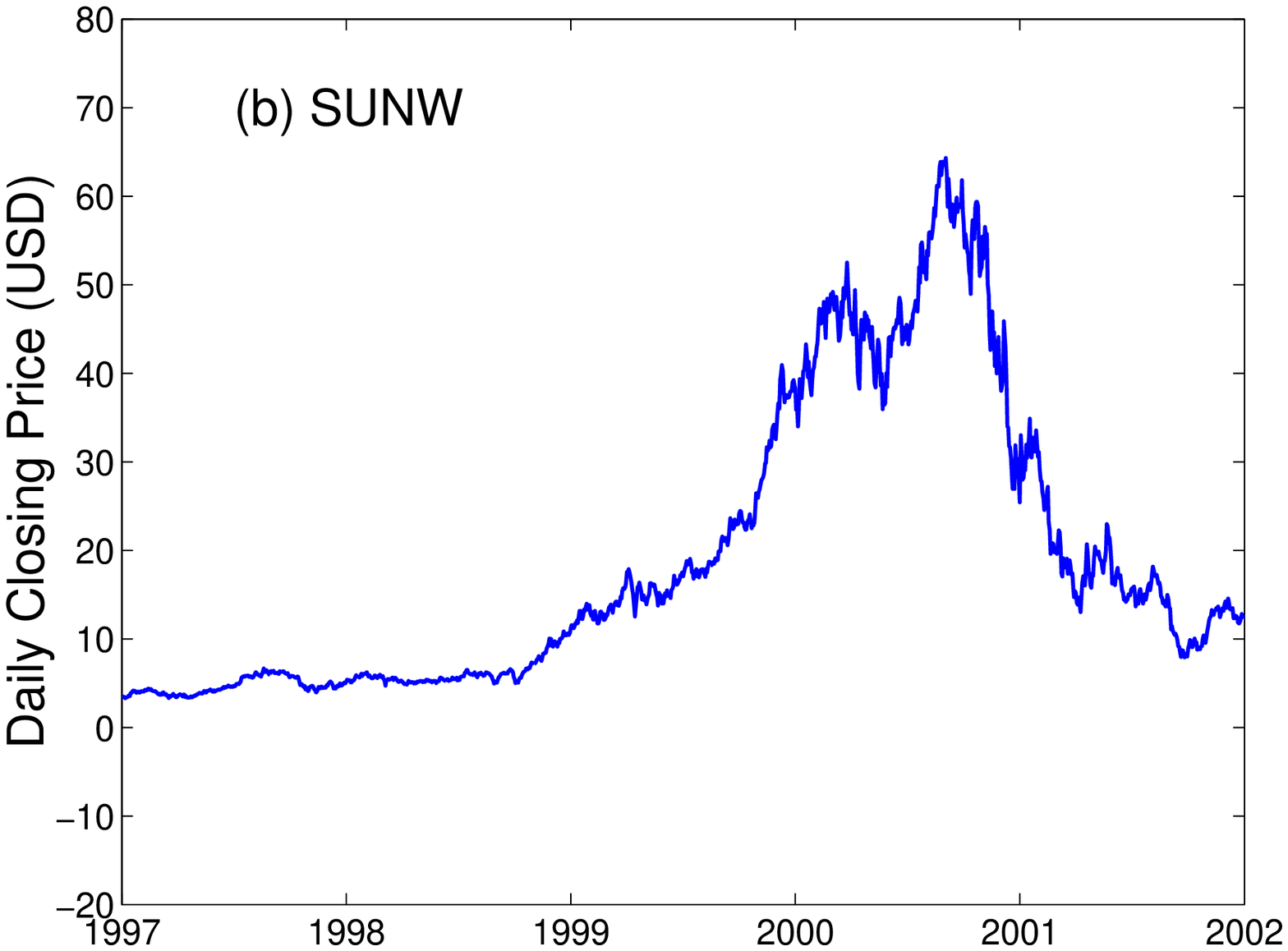} \hfill
\includegraphics[width=.30\textwidth]{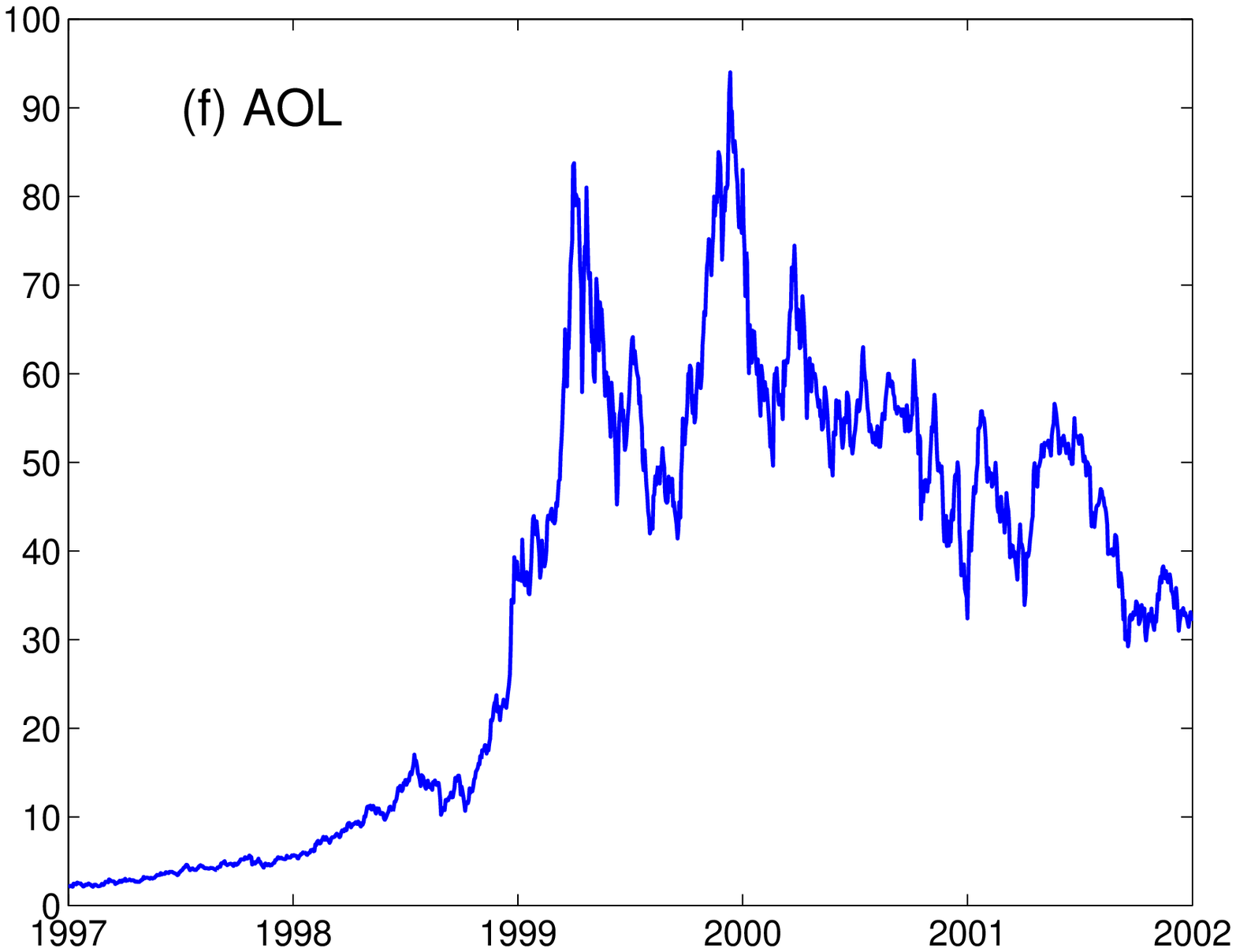} \hfill
\includegraphics[width=.30\textwidth]{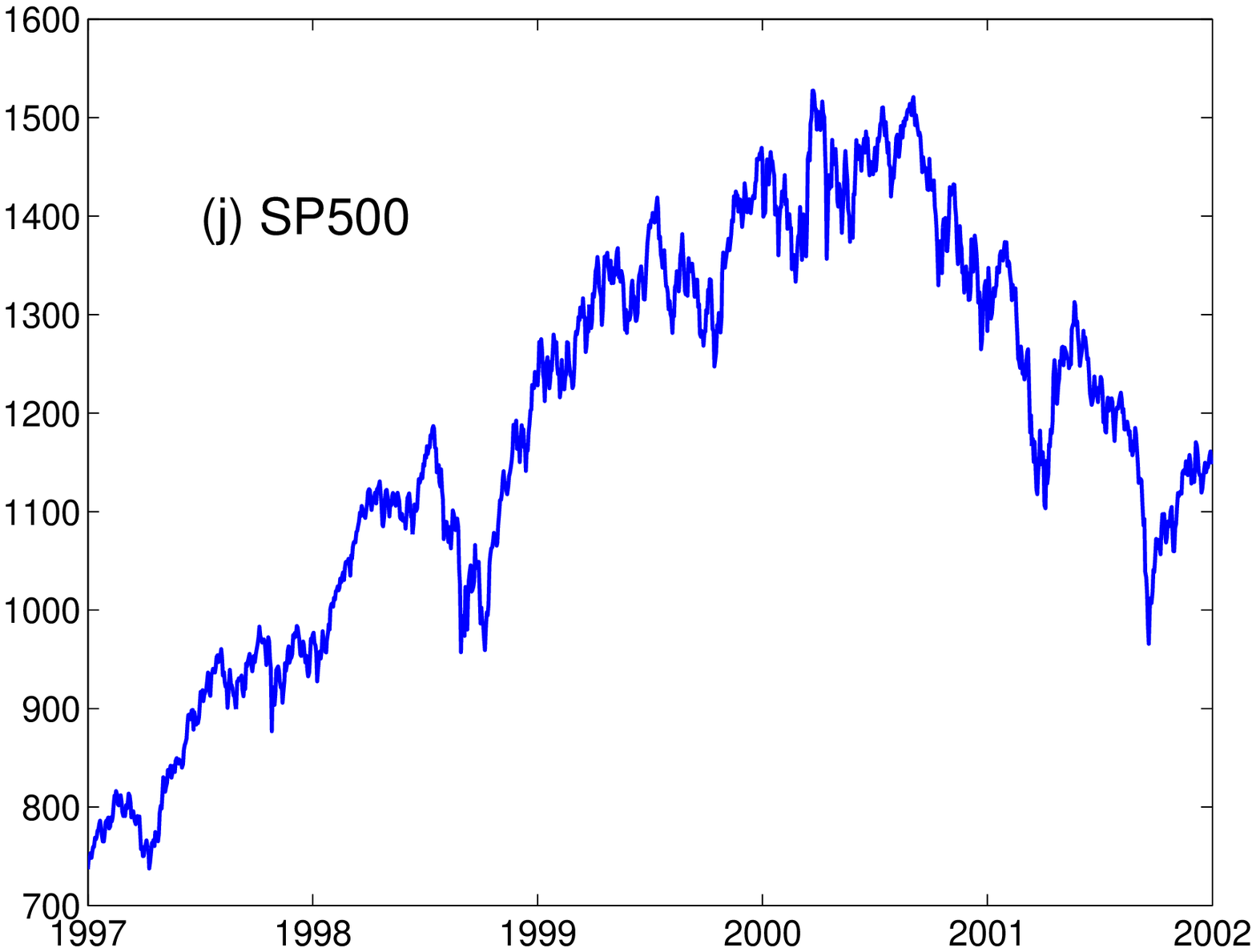} \vfill
\includegraphics[width=.30\textwidth]{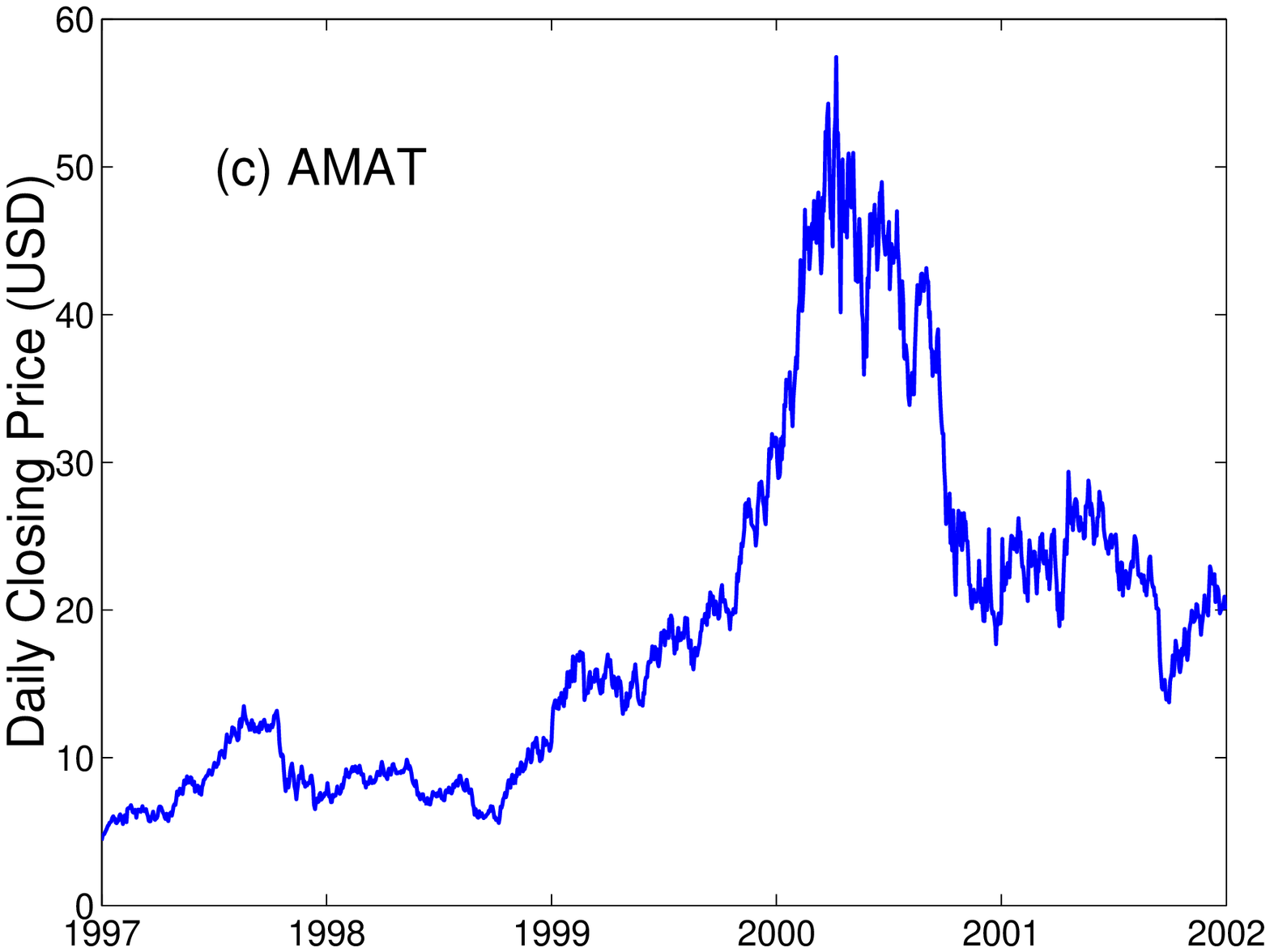} \hfill
\includegraphics[width=.30\textwidth]{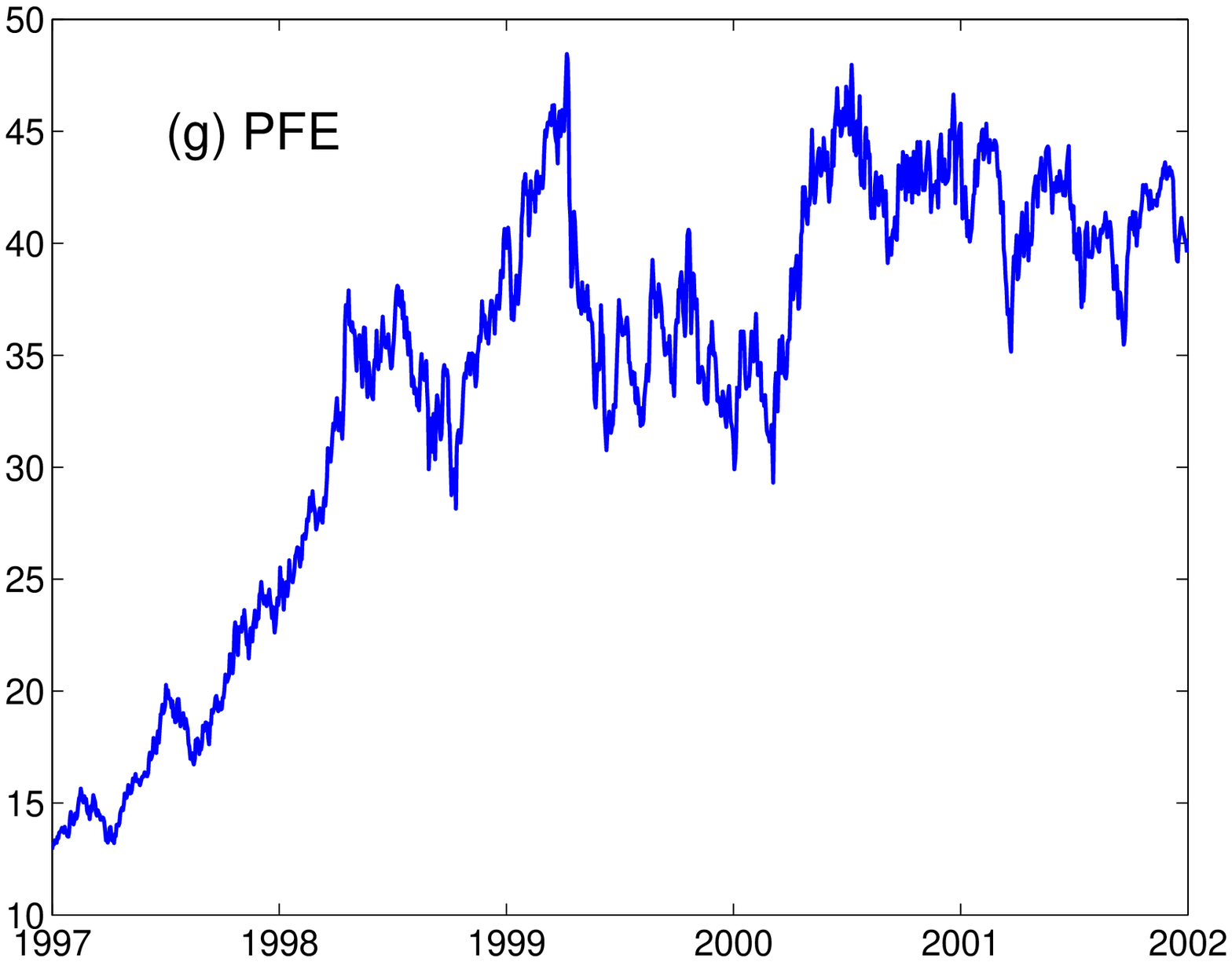} \hfill
\includegraphics[width=.30\textwidth]{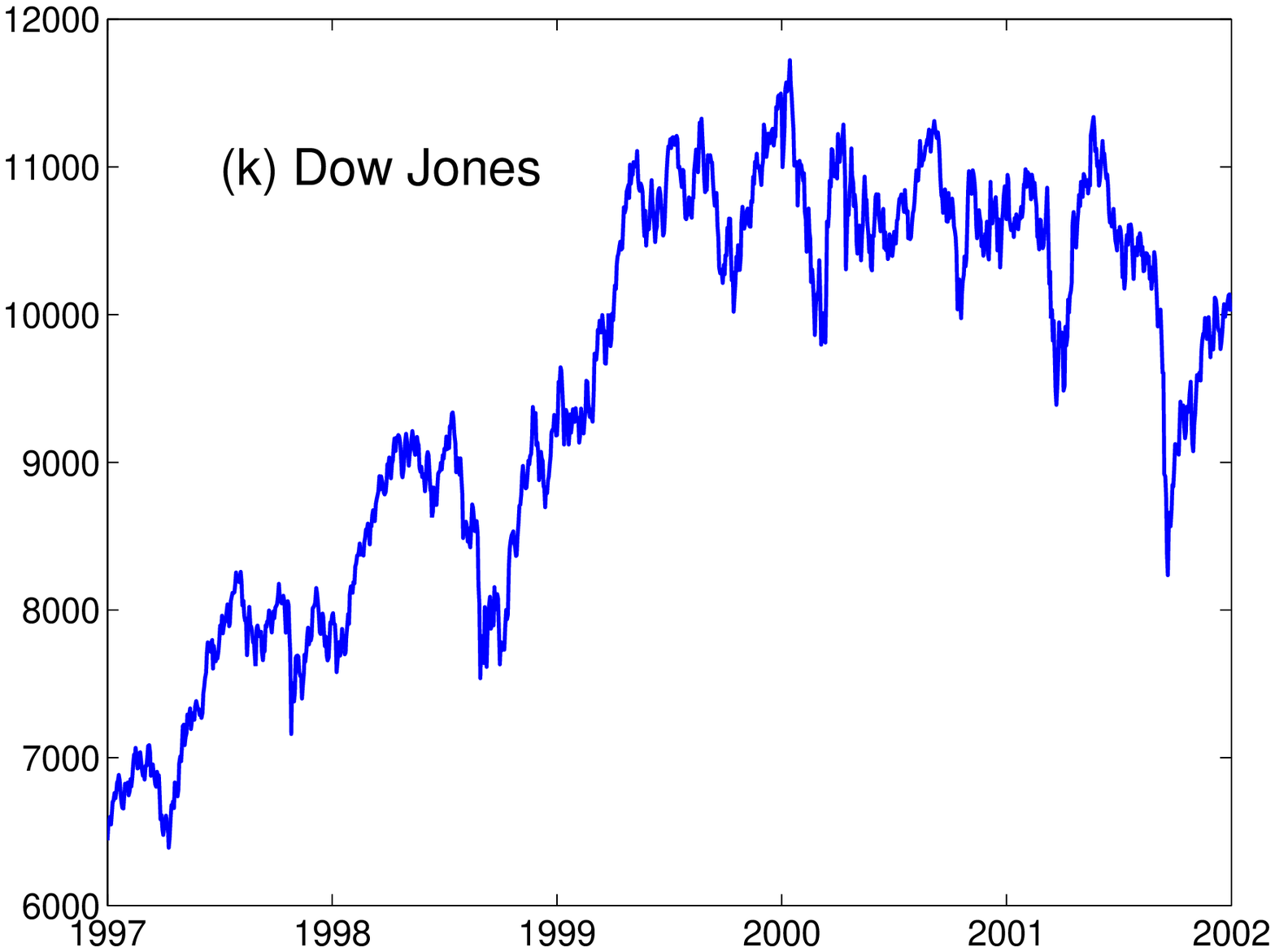} \vfill
\includegraphics[width=.30\textwidth]{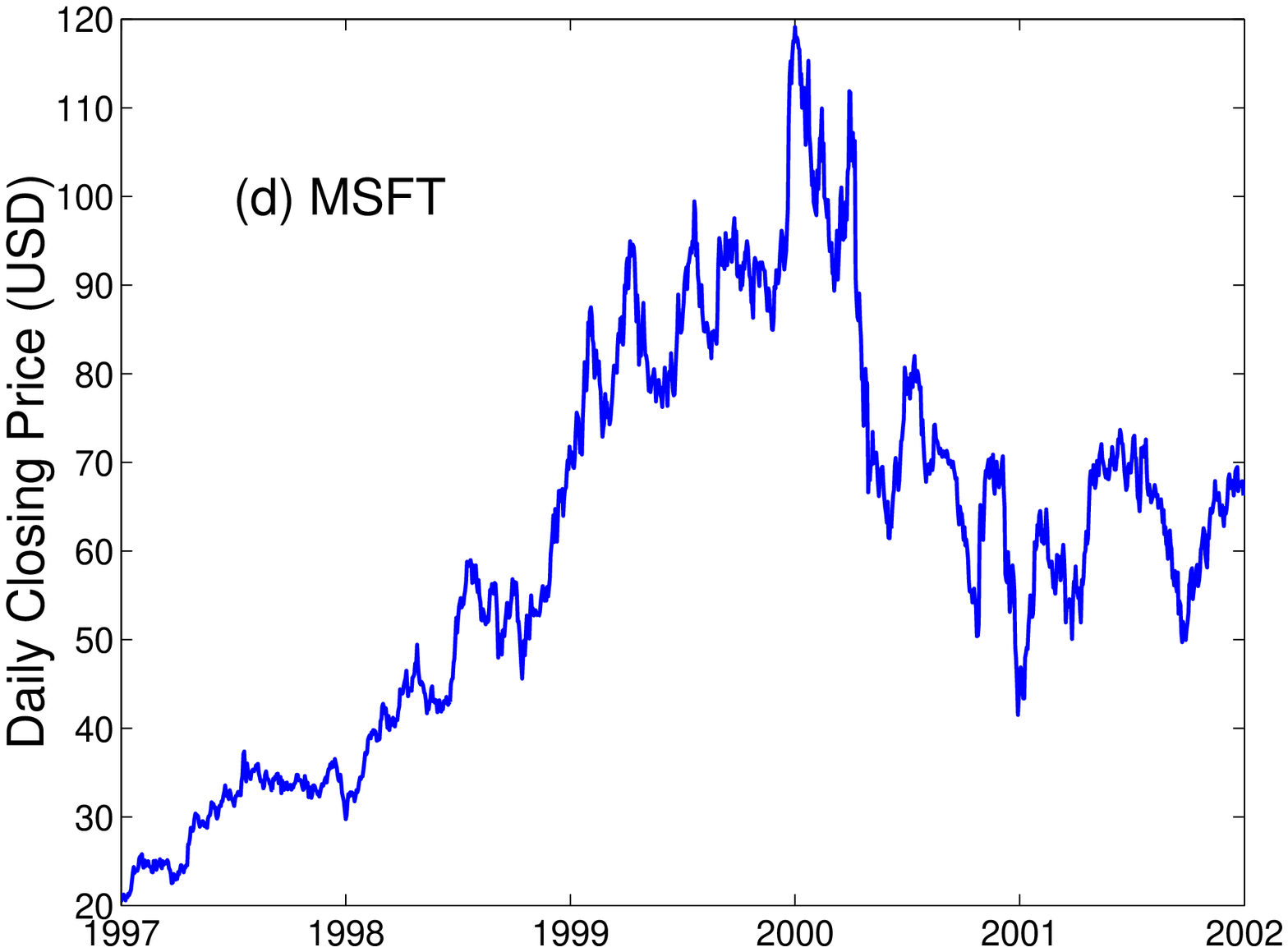} \hfill
\includegraphics[width=.30\textwidth]{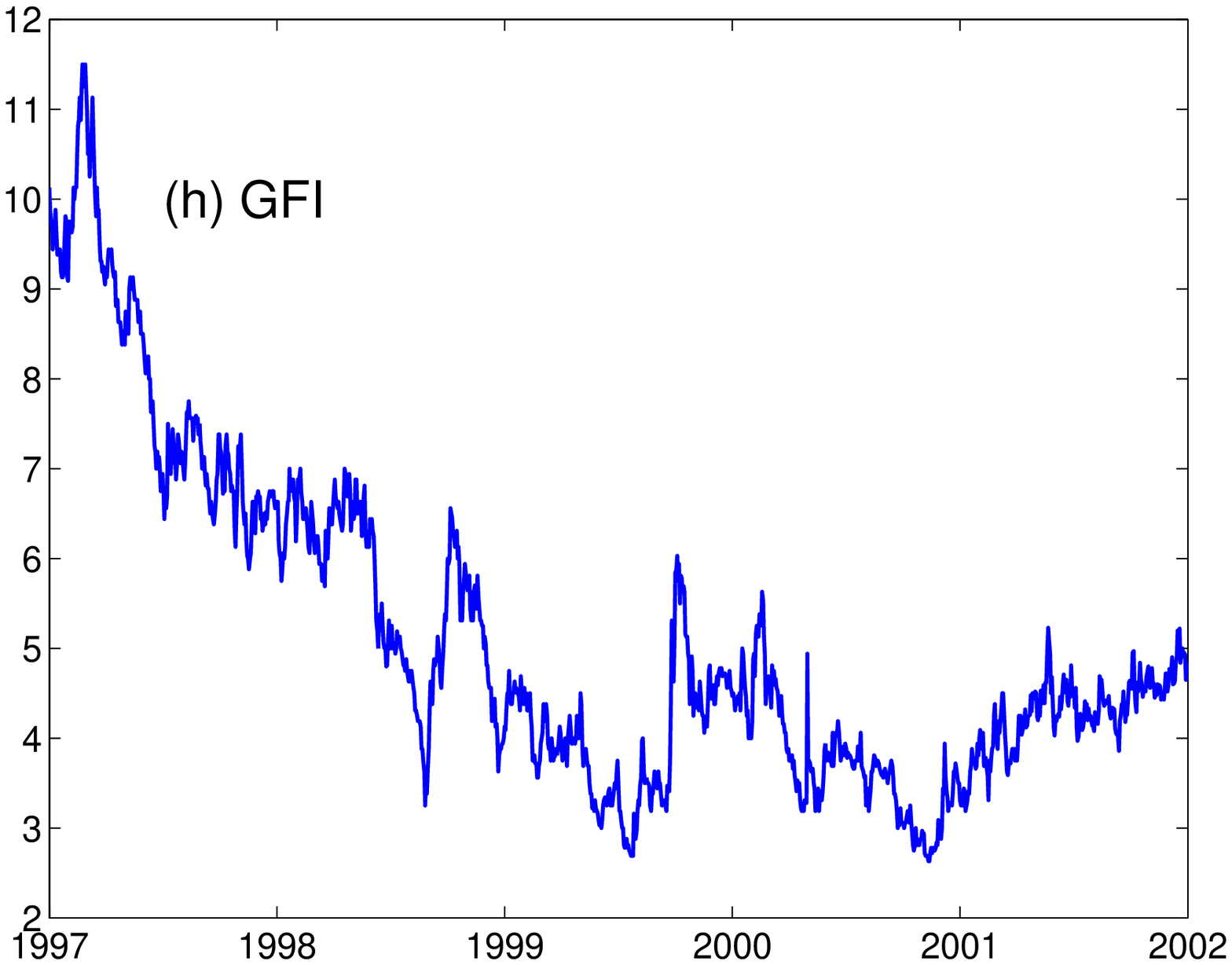} \hfill
\includegraphics[width=.30\textwidth]{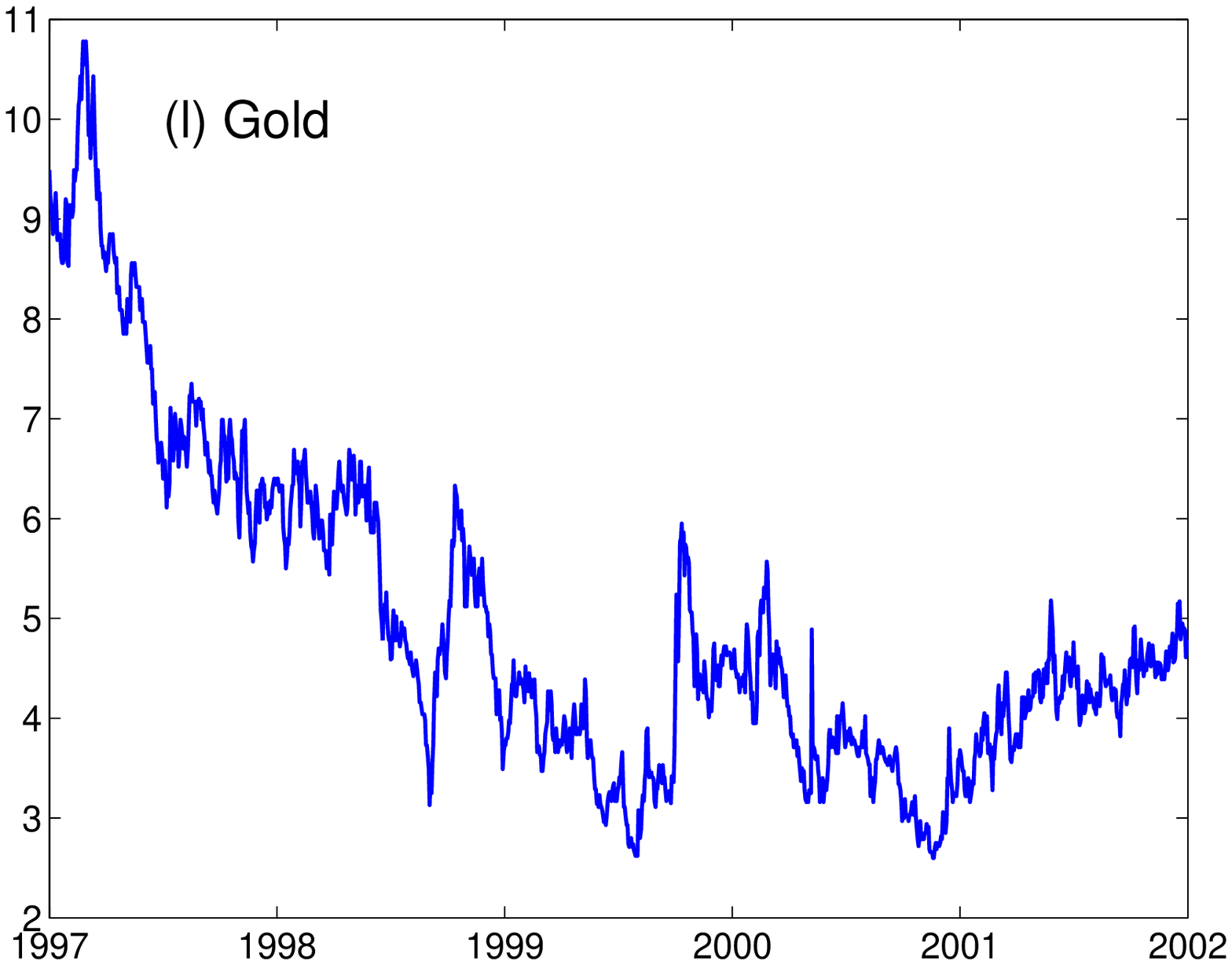} 
\caption{Daily closing
price $y(t)$ for the period Jan. 1, 1997 to Dec. 31, 2001 of stocks traded on
NASDAQ -- (a) CSCO, (b) SUNW, (c) AMAT, (d) MSFT, or traded on NYSE -- (e) GE,
(f) AOL, (g) PFE, (h) GFI;  three major financial indices -- (i) NASDAQ, (j)
S\&P500, (k) DJIA; the price of Gold (l)}
\label{eps1}
\end{figure}

\section{Classical Technical Analysis}
\label{sec:2}

Technical indicators like the  {\it moving average} and the $momentum$ are part
of the classical technical analysis and much used in efforts to predict market
movements \cite{Achelis}. One question is whether these techniques provide
adequate ways to read the trends.

Consider a time series $y(t)$ given at $N$ discrete times $t$. Let us recall that
the series (or signal) moving average $M_{\tau}(t)$ over a time interval $\tau$
is defined by

\begin{equation} M_{\tau}(t)=\frac{1}{\tau}\sum_{i=t}^{t+\tau-1} y(i-\tau) \qquad
t=\tau+1,\dots,N \end{equation} i.e. the average of $y$ over the last $\tau$ data
points. One can easily show that if the signal $y(t)$ increases (decreases) with
time, $M_{\tau}(t)<y(t)$ ($M_{\tau}(t)>y(t)$). Thus, the moving average captures
the trend of the signal given the period of time $\tau$. The intersections of the
price signal with a moving average can define so-called lines of resistance or
support \cite{Achelis}. The intersections between two moving averages, the
so-called ''death cross'' and ''gold cross'' in empirical finance \cite{Achelis},
are usually used to identify points of drastic changes in the trend of the
signal. The {\it cross density} characterizes the signal roughness \cite{nvmama}.

The so called momentum indicator \cite{Achelis} is another instrument of the
technical analysis and we will refer to it here as the {\it classical momentum
indicator} (CMI). The classical momentum indicator of a stock is a {\it moving
average of the momentum} defined over a time interval $\tau$ as
\begin{equation} R^{\Sigma}_{\tau}(t) = \sum_{i=t}^{t+\tau-1}
\frac{y(i)-y(i-\tau)}{\tau} \qquad t=\tau+1,\dots,N \end{equation}
The classical momentum indicator (CMI) $R^{\Sigma}_{\tau}$ for time interval,
$\tau=21$, i.e. one month is shown in Fig. 2(a) for GE.

\section{Generalized Technical Analysis}

Stock markets do have another component beside prices or volatilities. This is
the volume of transactions (Fig. 2 (a) for GE) which we have considered as the
''physical mass'' of stocks, in a generalized technical analysis scheme
\cite{epjb_gta}. Remember that the number of shares is constant over rather long
time intervals, usually like the mass of an object.
\begin{figure} \centering \includegraphics[width=.48\textwidth]{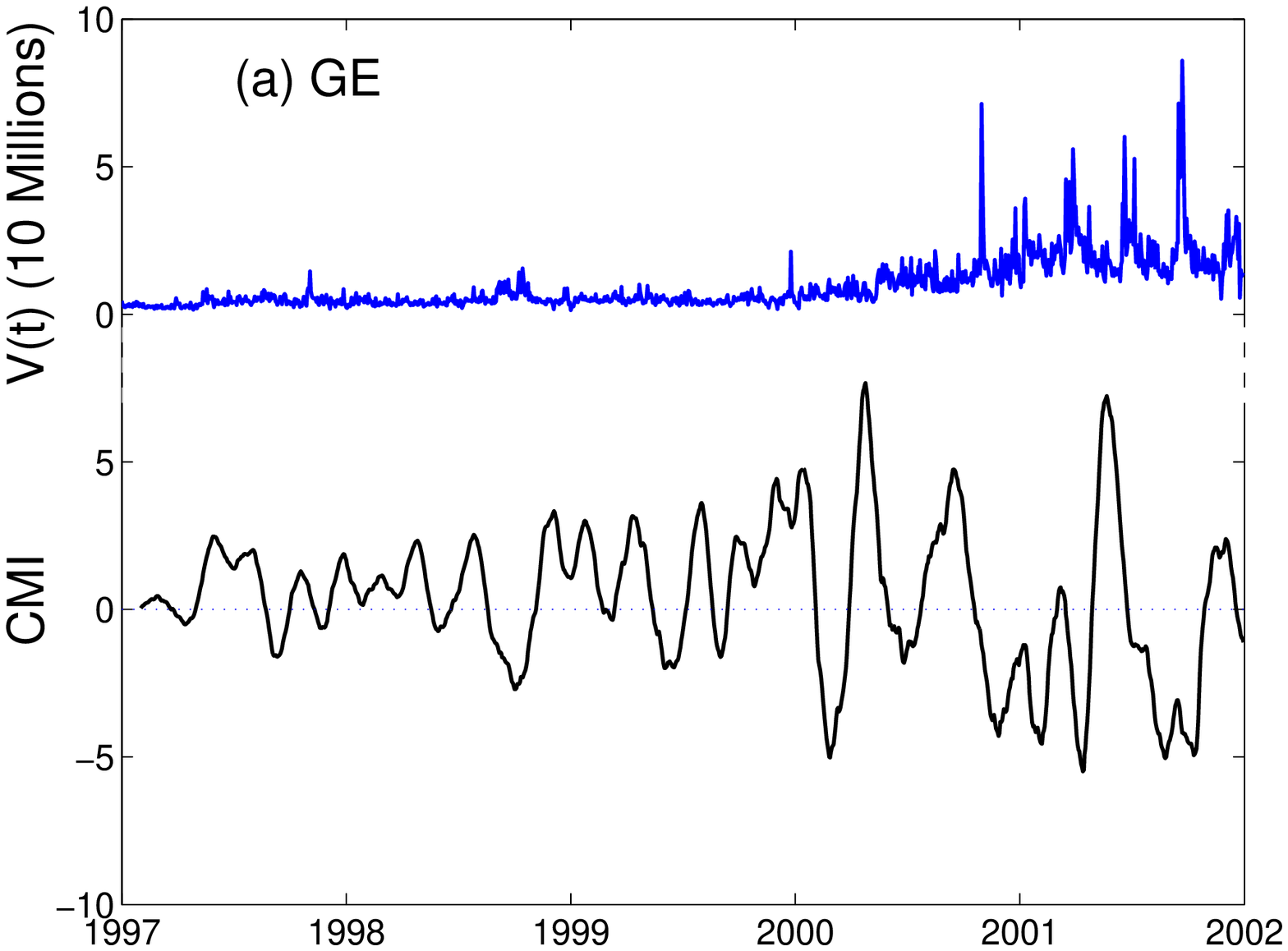}
\hfill \includegraphics[width=.48\textwidth]{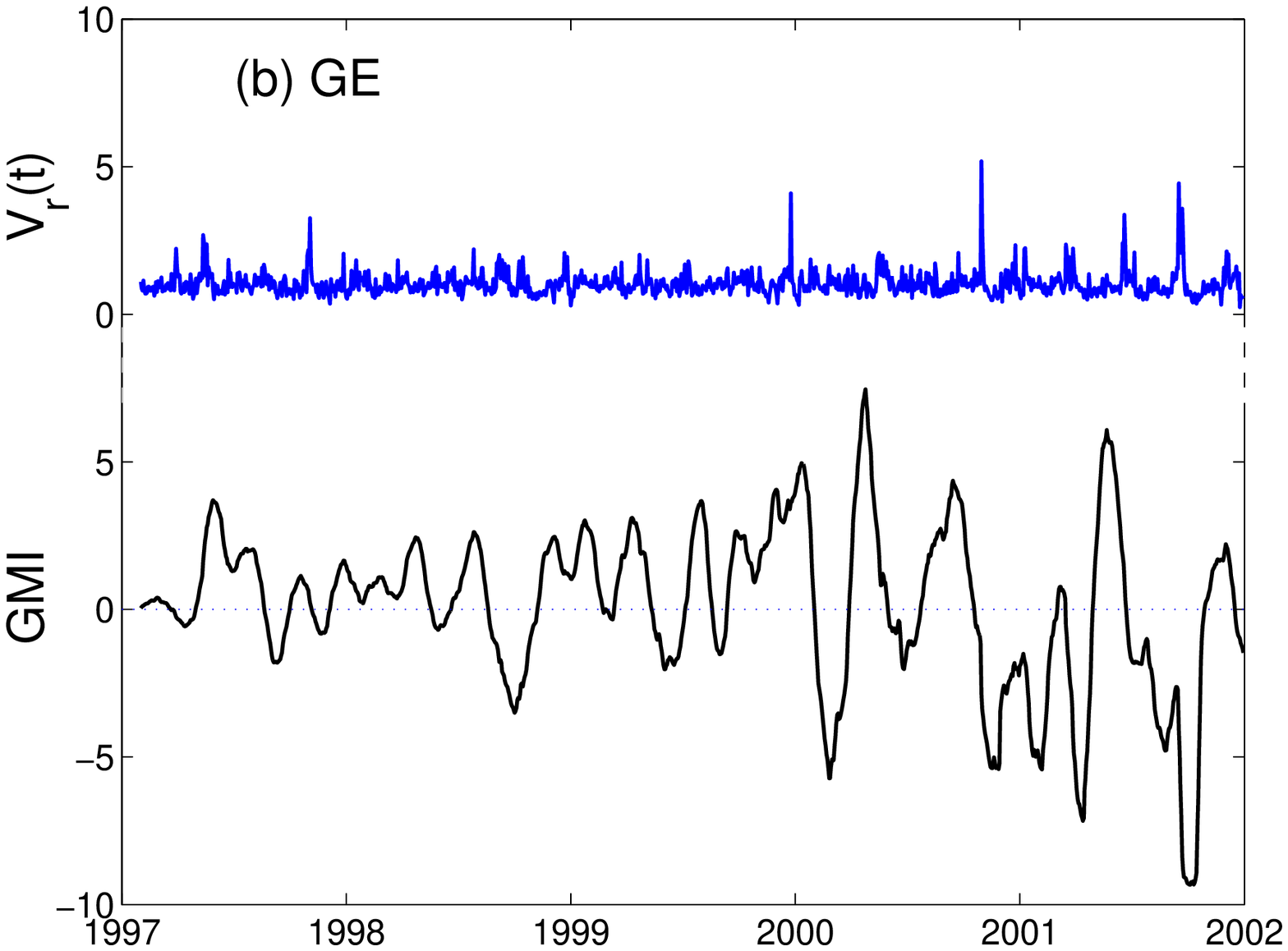} \caption{(a) Volume of
transactions $V(t)$ (in 10 Millions) and classical momentum indicator (CMI)
$R^{\Sigma}_{\tau}(t)$ for $\tau$=21~days, Eq. (2), for General Electric (GE) for
the period Jan. 1, 1997 to Dec. 31, 2001; (b) Reduced volume of transactions
$V_r(t)$ and generalized momentum indicator (GMI) $\widetilde
R^{\Sigma}_{\tau}(t)$ for $\tau$=21~days, Eq. (3), for General Electric (GE) for
the period Jan. 1, 1997 to Dec. 31, 2001} \label{eps2} \end{figure}

\begin{figure} \centering
\includegraphics[width=.30\textwidth]{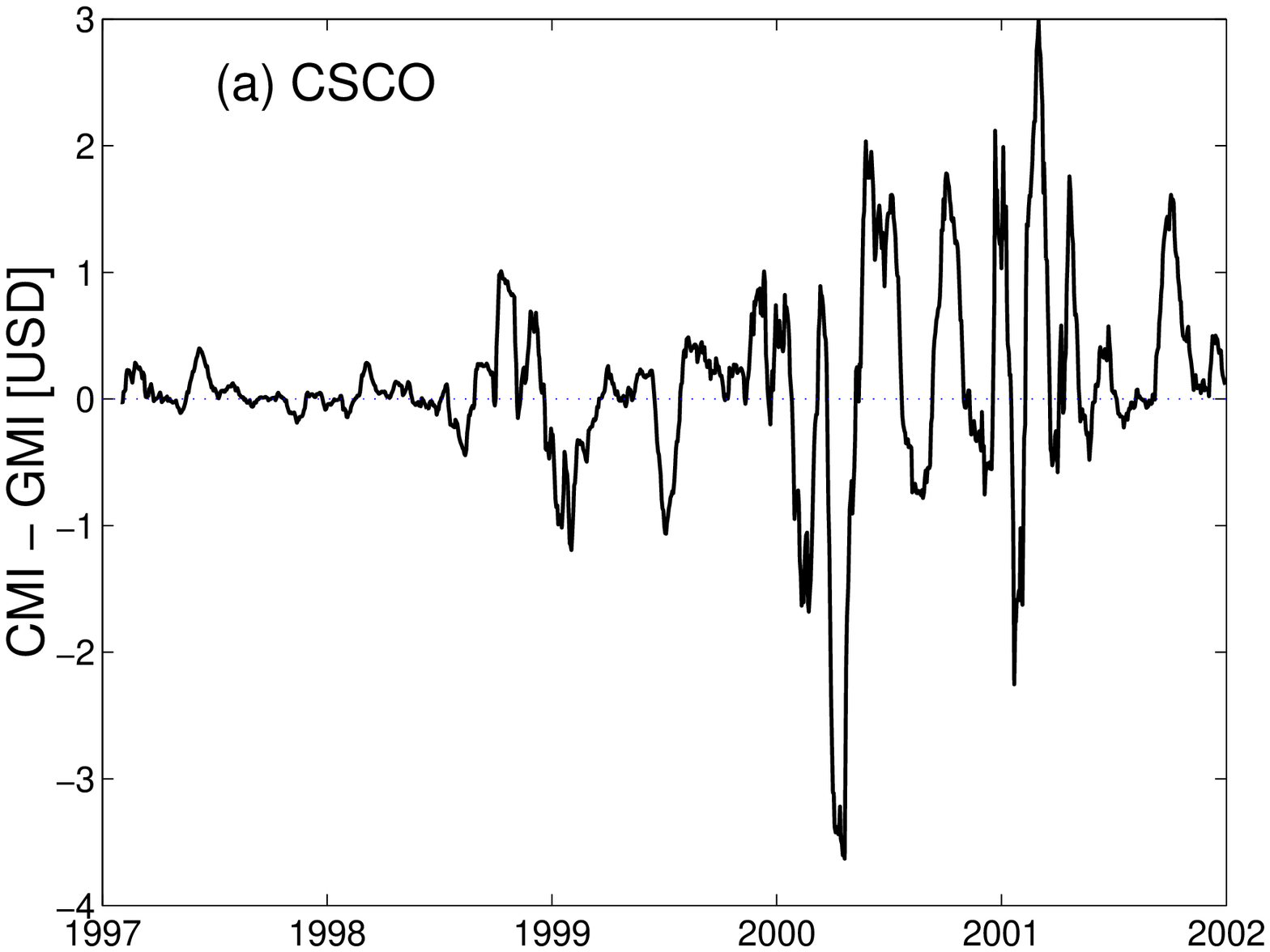} \hfill
\includegraphics[width=.30\textwidth]{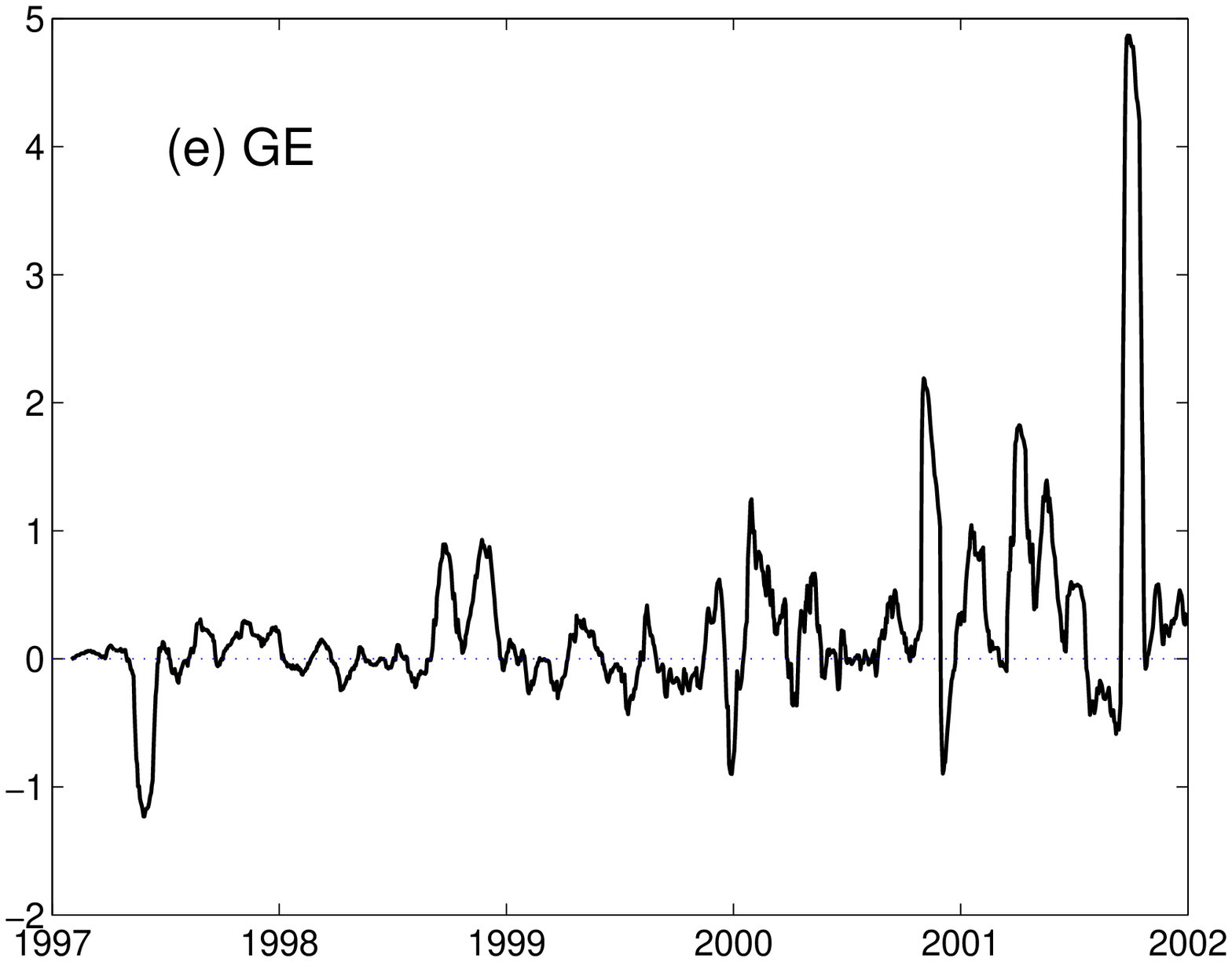} \hfill
\includegraphics[width=.30\textwidth]{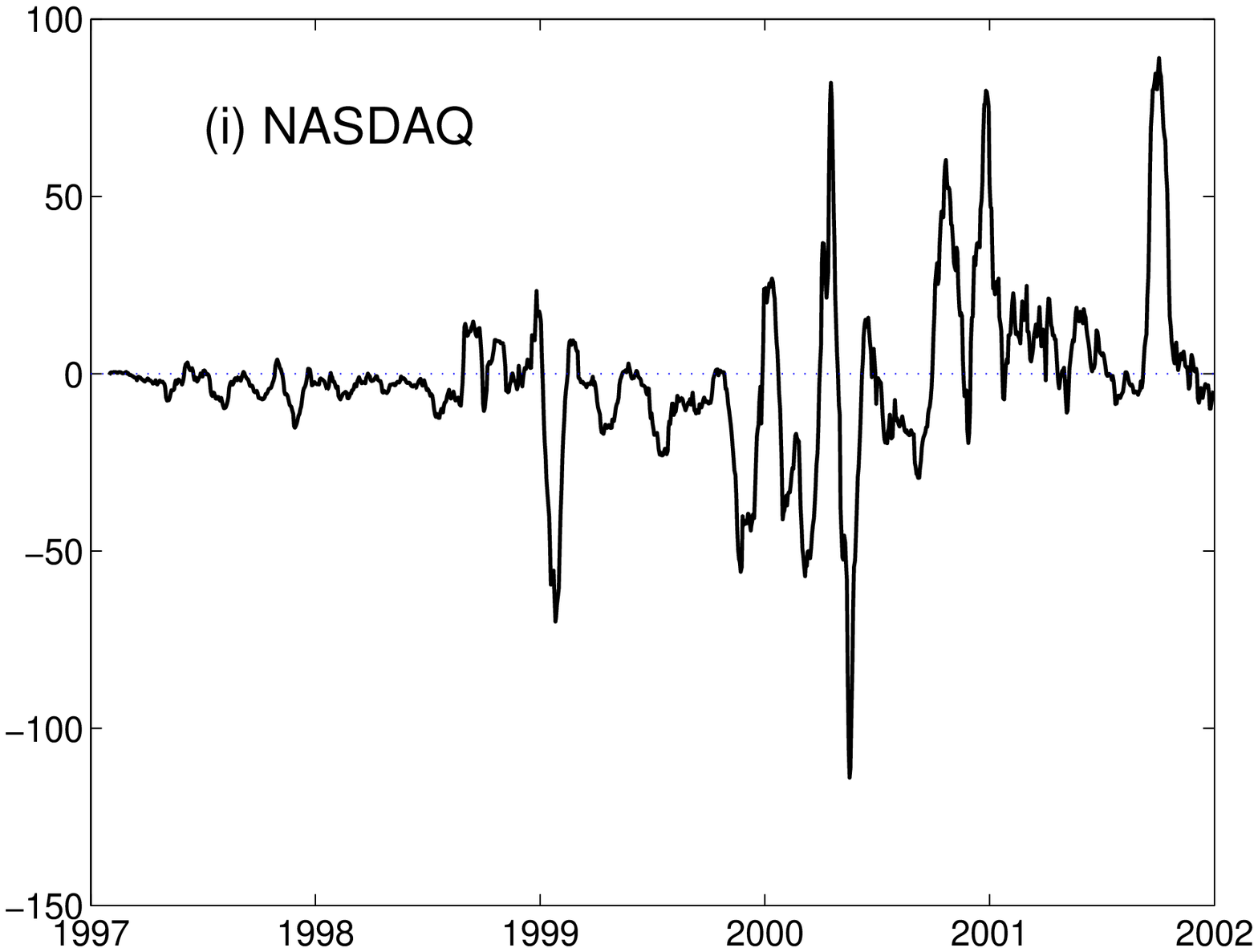} \vfill
\includegraphics[width=.30\textwidth]{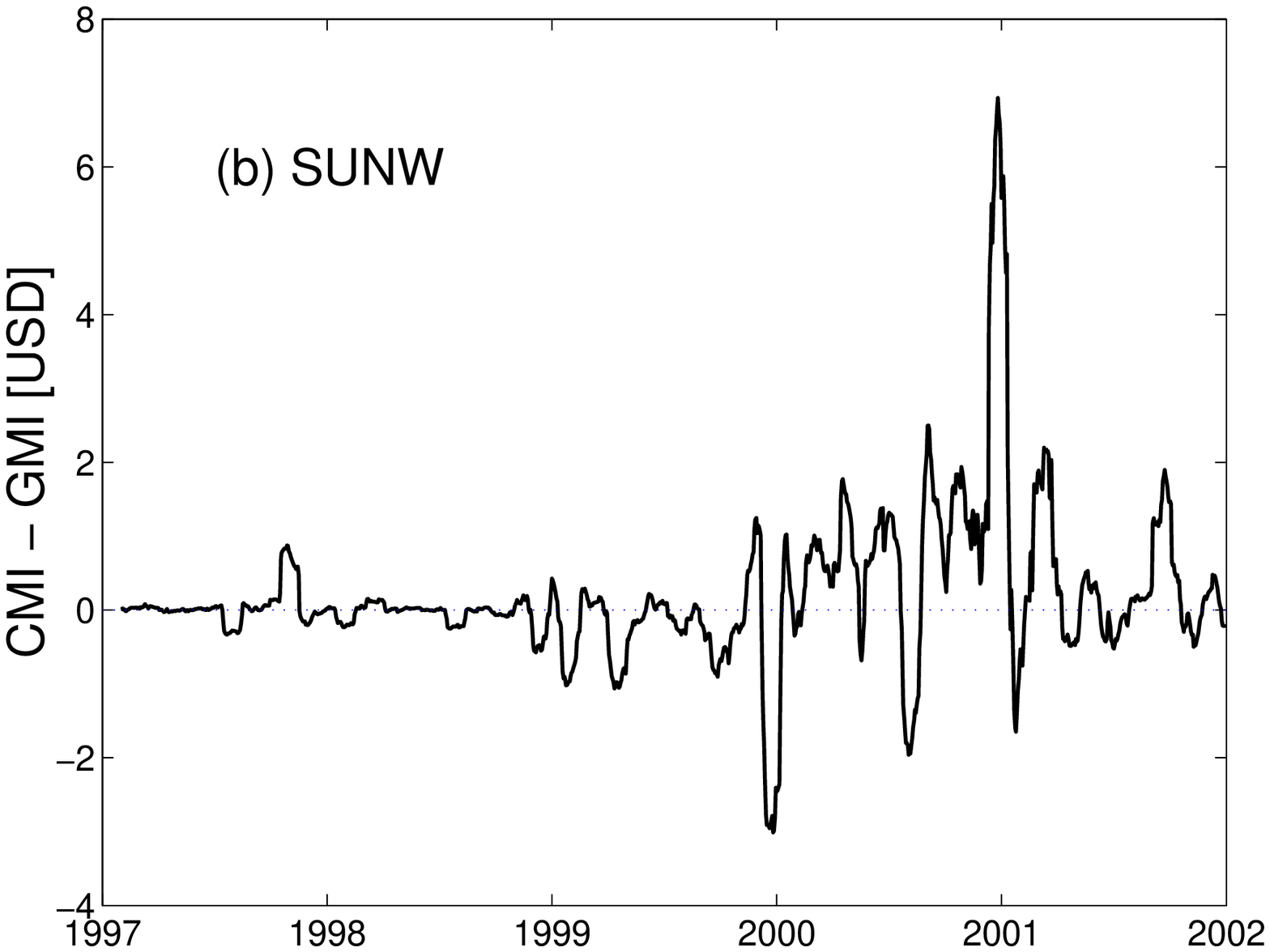} \hfill
\includegraphics[width=.30\textwidth]{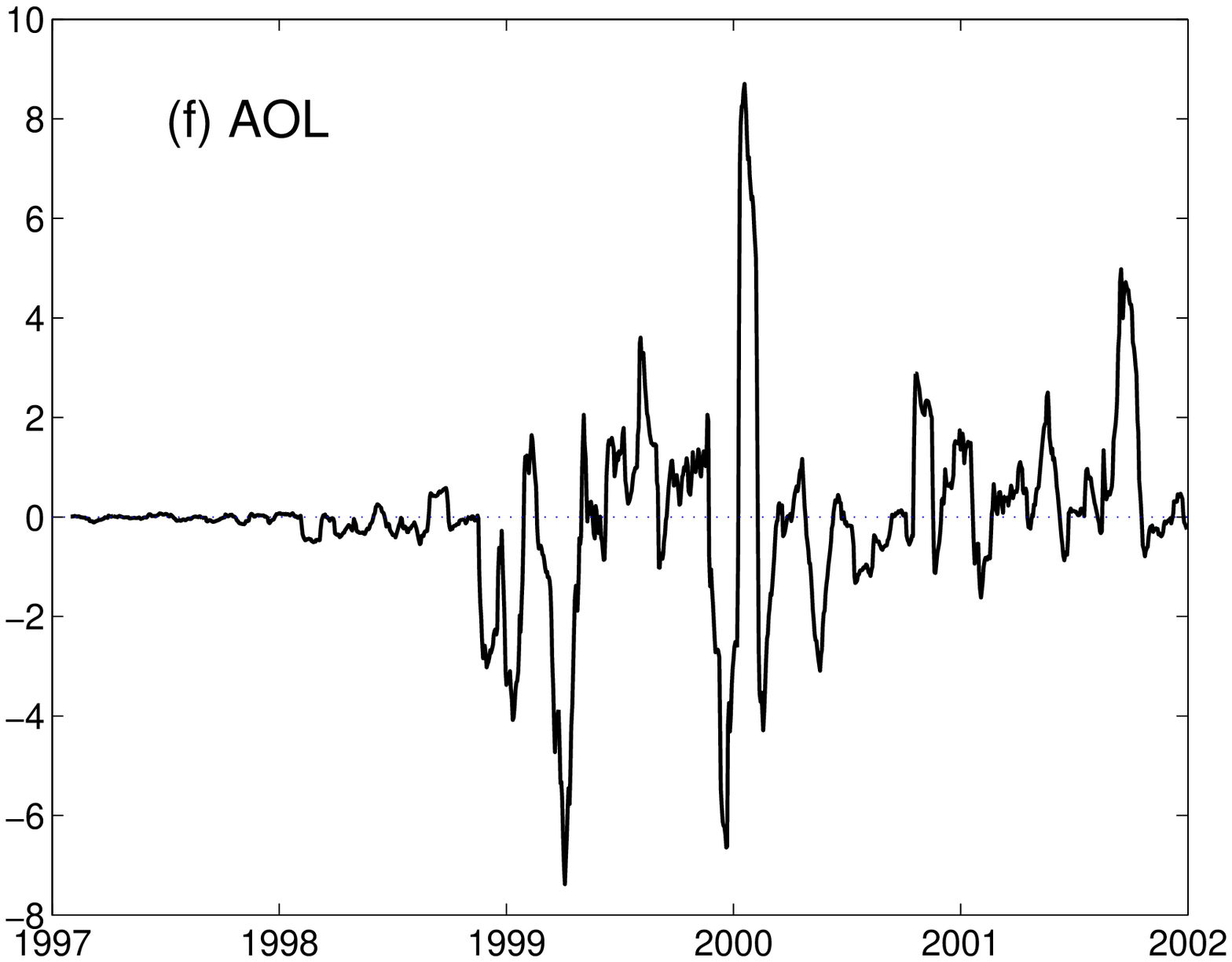} \hfill
\includegraphics[width=.30\textwidth]{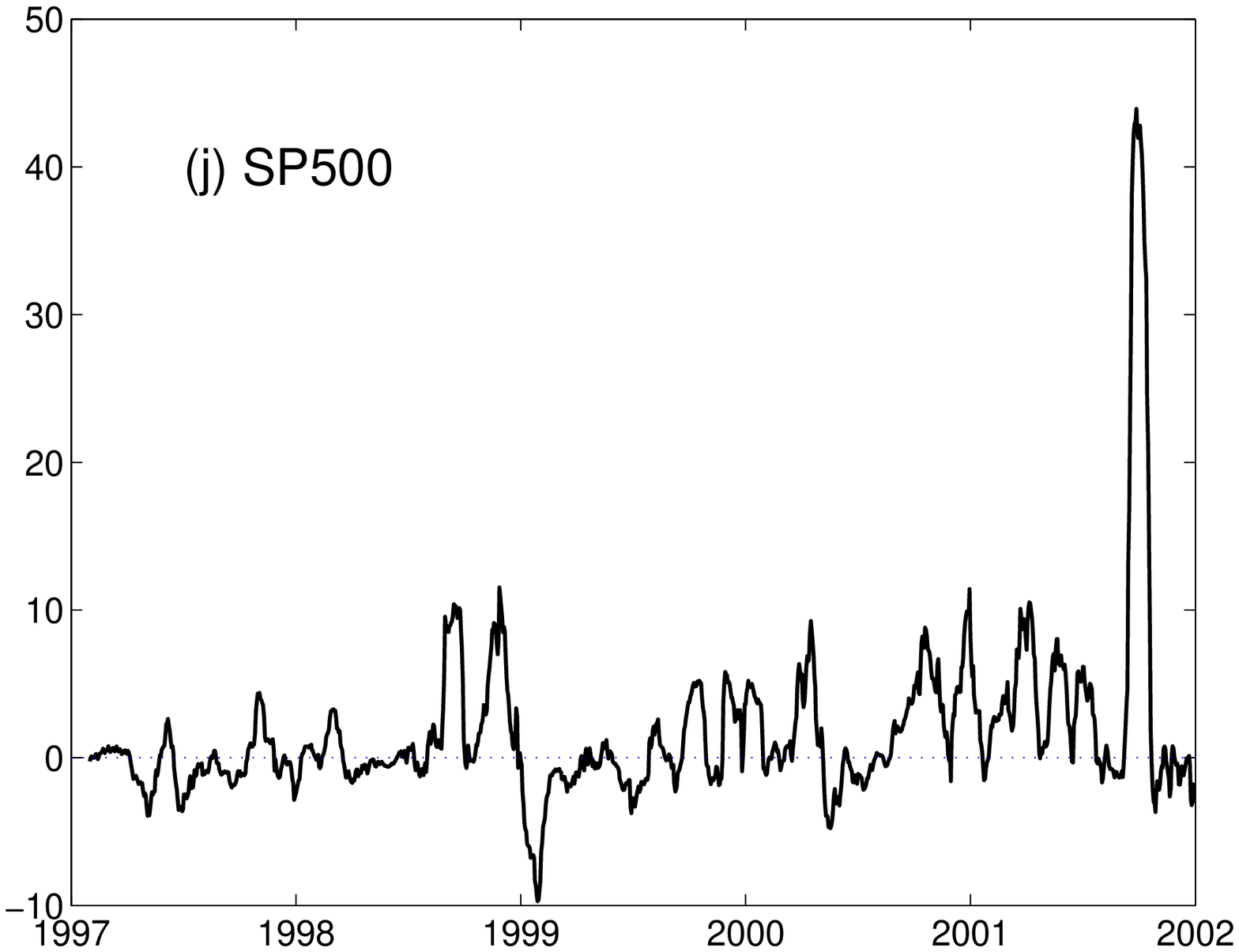} \vfill
\includegraphics[width=.30\textwidth]{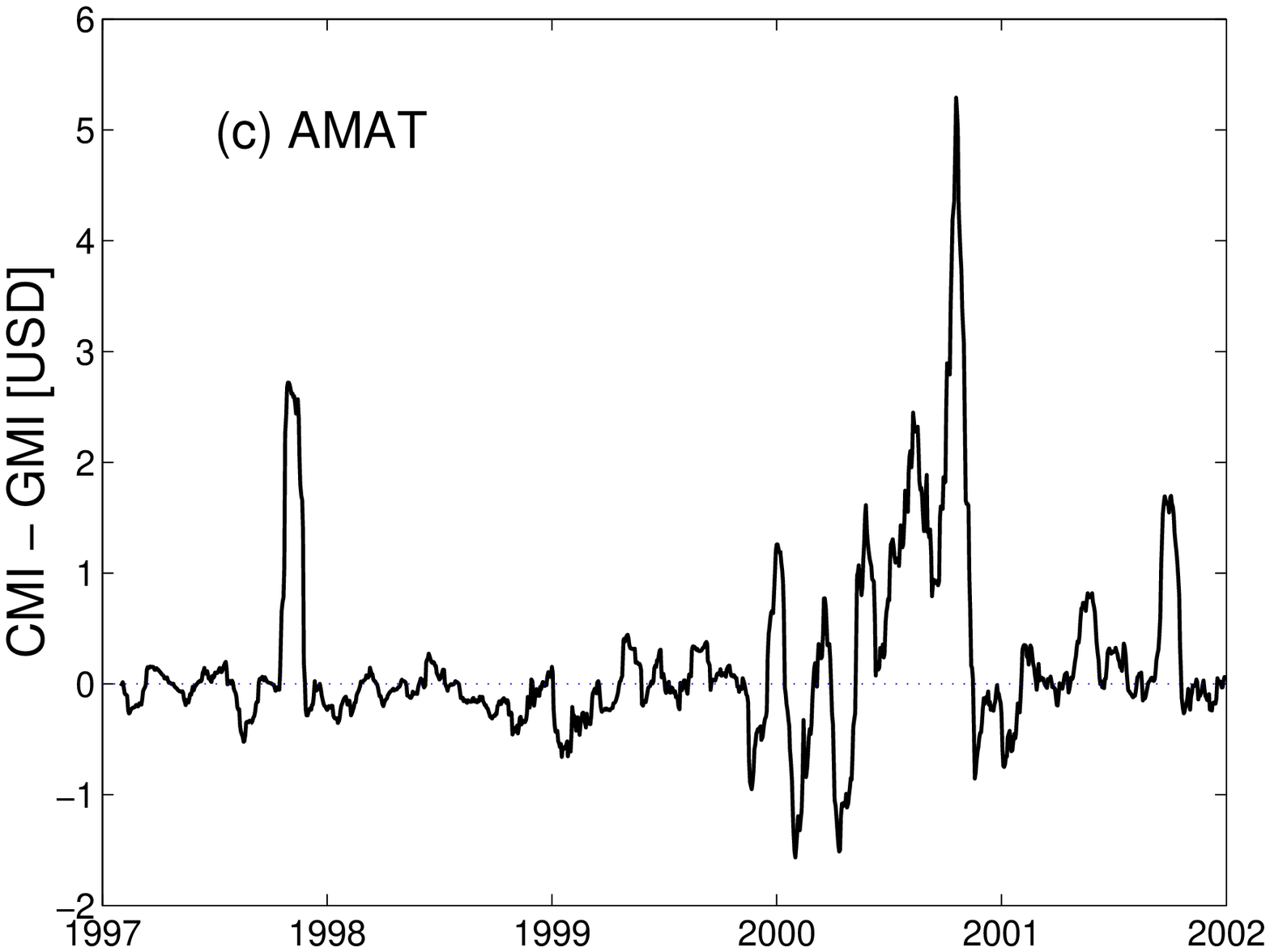} \hfill
\includegraphics[width=.30\textwidth]{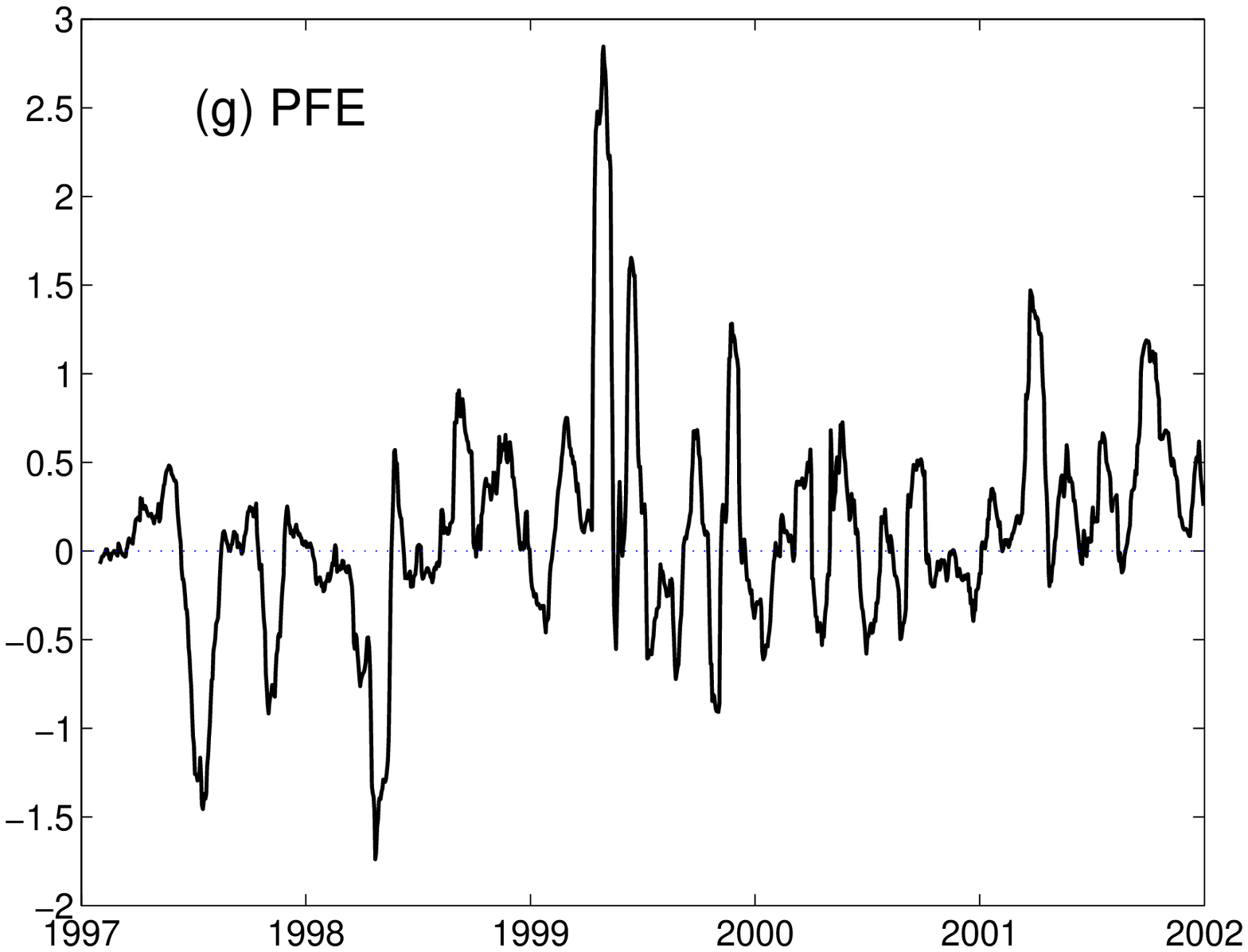} \hfill
\includegraphics[width=.30\textwidth]{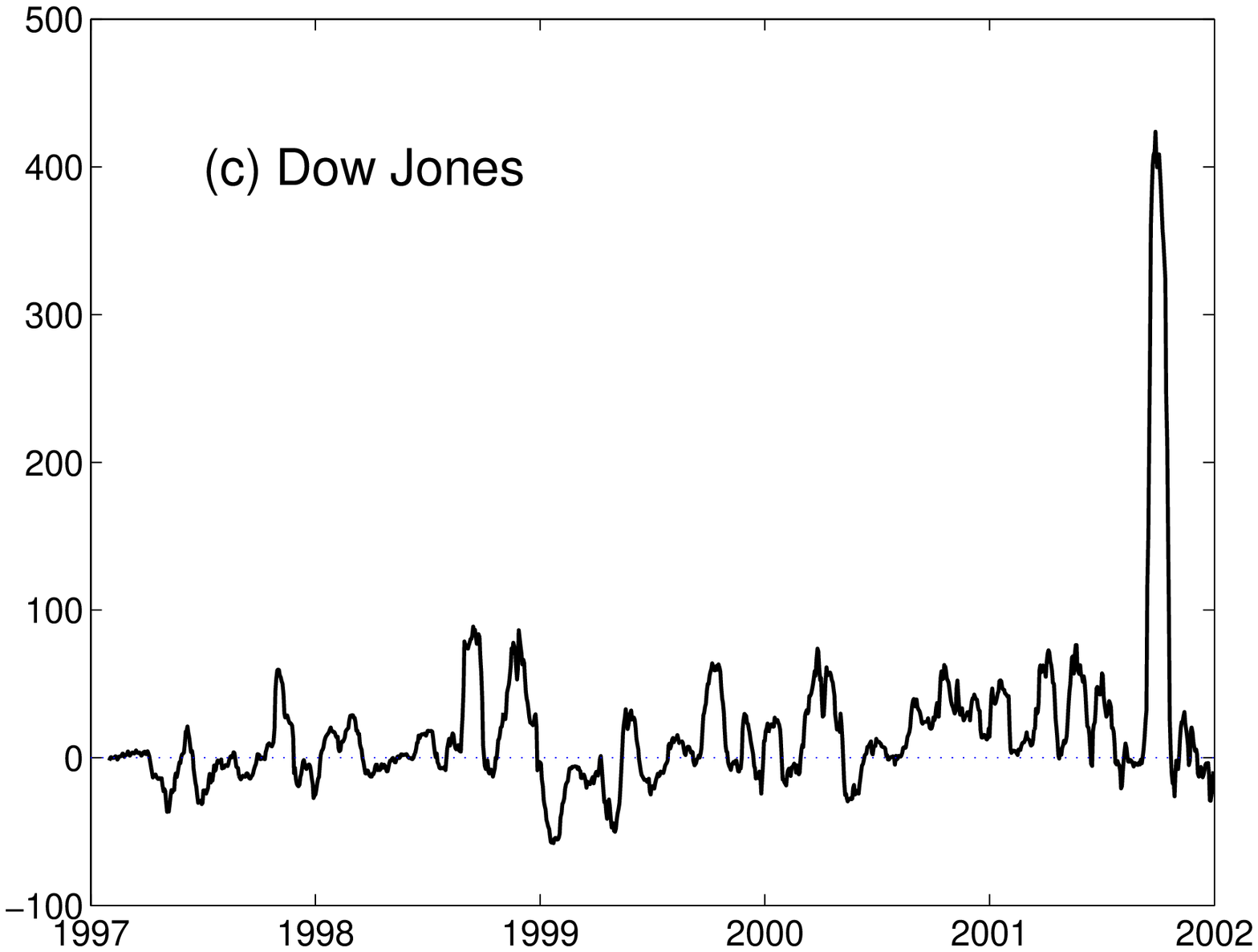} \vfill
\includegraphics[width=.30\textwidth]{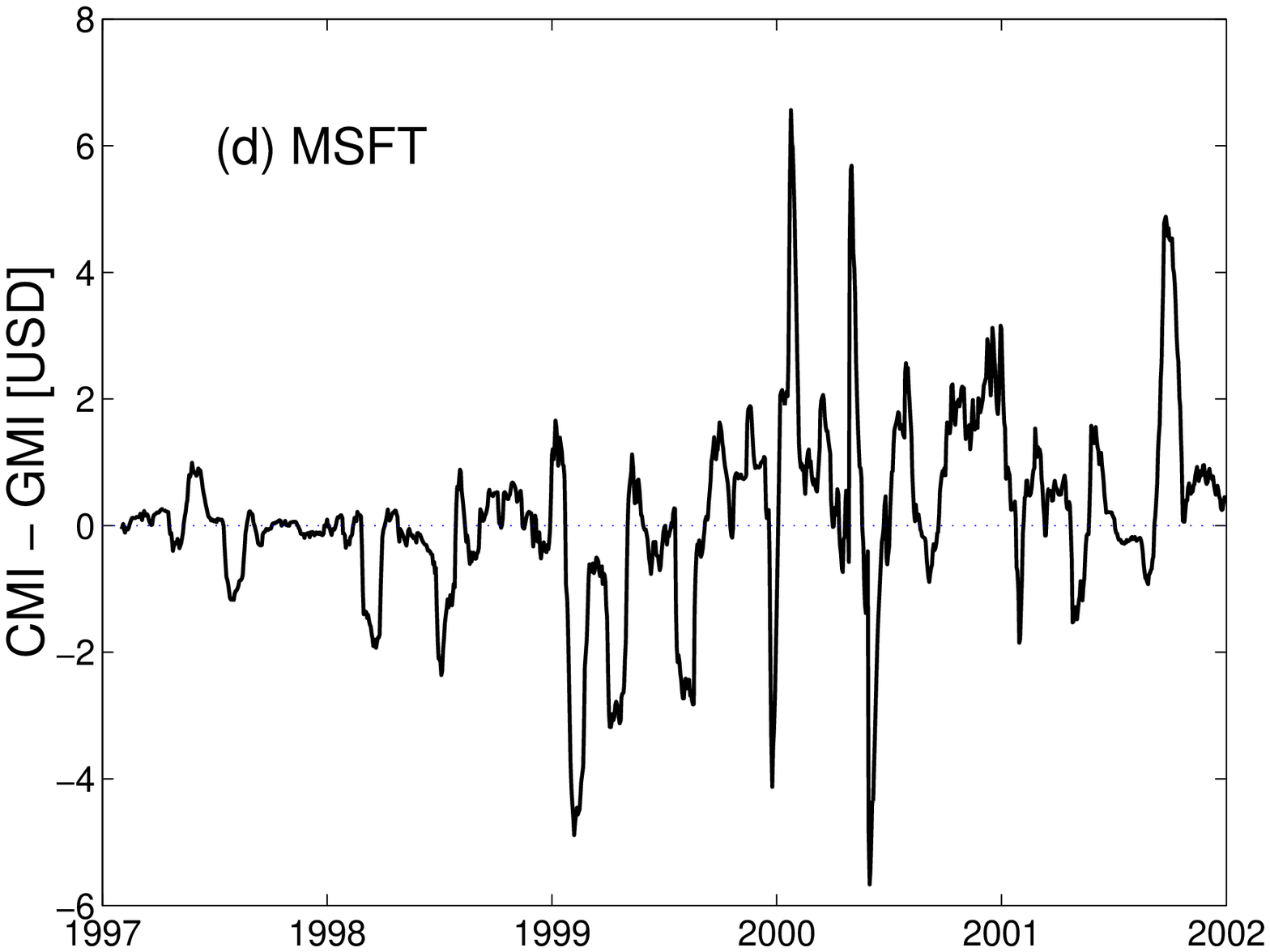} \hfill
\includegraphics[width=.30\textwidth]{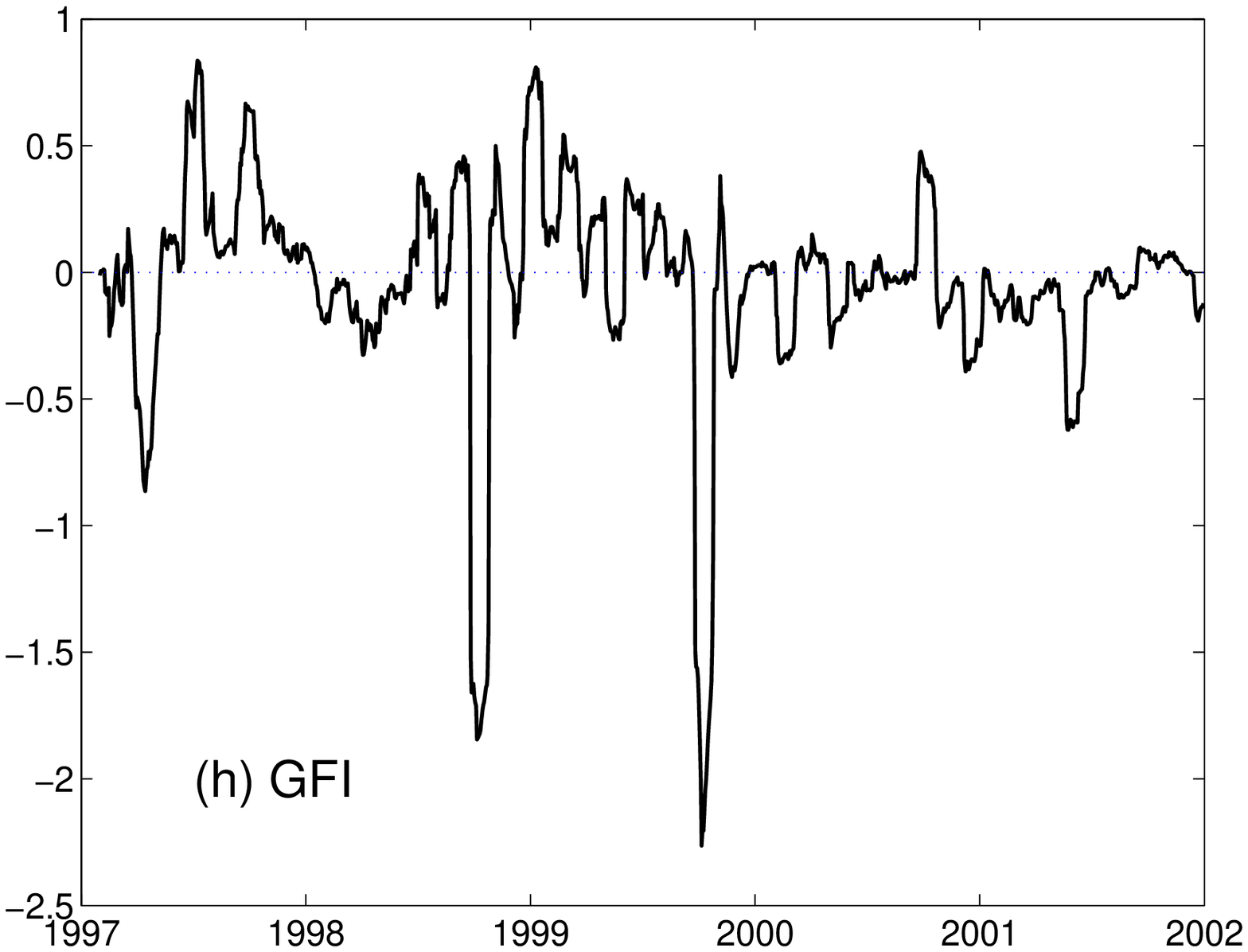} \hfill
\includegraphics[width=.30\textwidth]{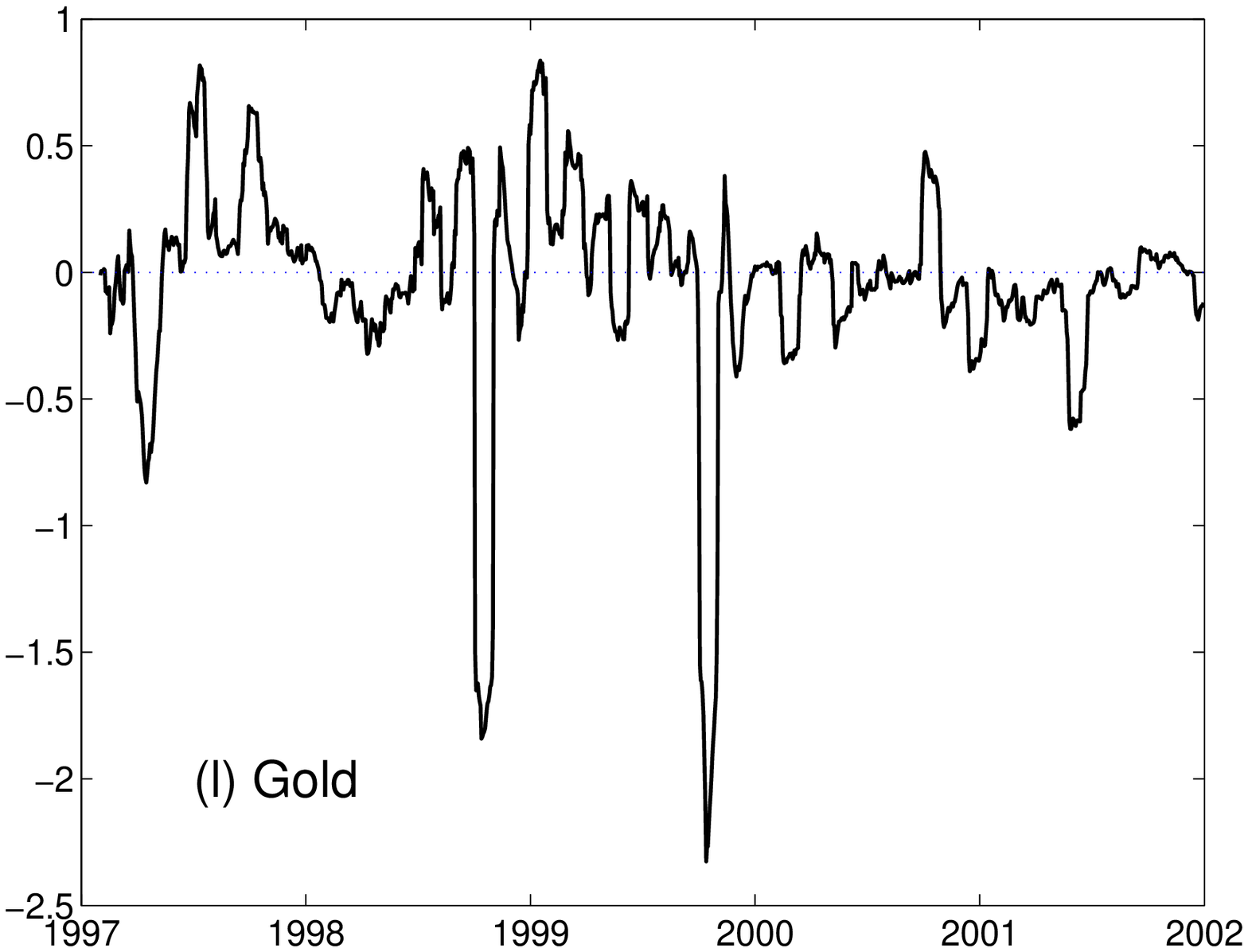} 
\caption{Difference between classical momentum indicator (CMI) 
$R^{\Sigma}_{\tau}(t)$ and generalized momentum indicator (GMI) 
$\widetilde R^{\Sigma}_{\tau}(t)$ with $\tau$=21~days from 
Jan. 1, 1997 to Dec. 31, 2001 for stocks traded on NASDAQ -- 
(a) CSCO, (b) SUNW, (c) AMAT, (d) MSFT;  stocks traded on NYSE -- 
(e) GE, (f) AOL, (g) PFE, (h) GFI;  three financial indices -- 
(i) NASDAQ, (j) S\&P500, (k) DJIA, and Gold  (l)} \label{eps3}
\end{figure}

Consider $V(t)$ to be the volume of transactions of a stock with price $y(t)$ at
time $t$. A generalized momentum $\widetilde R_{\tau}$ over a time interval
$\tau$ can be
\begin{equation} \widetilde R_{\tau}(t)=\frac{V(t)}{<V(t)>_{\tau}}\cdot
\frac{y(t)-y(t-\tau)}{\tau}=m(t)\frac{\Delta x}{\Delta t}, \qquad
t=\tau+1,\dots,N \end{equation} where the total volume of transactions over the
interval $\tau$ is $<V>_{\tau}\tau=\sum_{i=1}^{\tau}V(i)$. In so doing, we
introduce some analogy to a {\it generalized time dependent mass} $m(t)$ that
contains some sort of history of the intercorrelations between price and volume
of transactions of a stock during the $\tau$ interval. The total volume in the
denominator is introduced for a normalization purpose. The transaction volume can
also be represented as a rescaled volume of transactions $V_r(t)$
\begin{equation}  V_r(t) = \frac{V(t)}{<V>_{\tau}} \end{equation} which is
plotted in Fig. 2 (b) for GE.

We further consider a {\it moving average of the generalized momentum} which is
called the generalized momentum indicator (GMI)
\begin{equation} \widetilde R^{\Sigma}_{\tau}(t)=\sum_{i=t}^{t+\tau-1}
\frac{V(t)}{<V>_{\tau}} \cdot \frac{y(i)-y(i-\tau)}{\tau} \qquad t=\tau+1,\dots,N
\end{equation} The difference CMI minus GMI for all data of interest is
plotted in Fig. 3.

\section{Investment strategy}

A simple investment strategy can be suggested based on the trends of the market
and using both the price per share and the volume of transactions incorporated in
the generalized momentum indicator.
\begin{figure} \centering
\includegraphics[width=.48\textwidth]{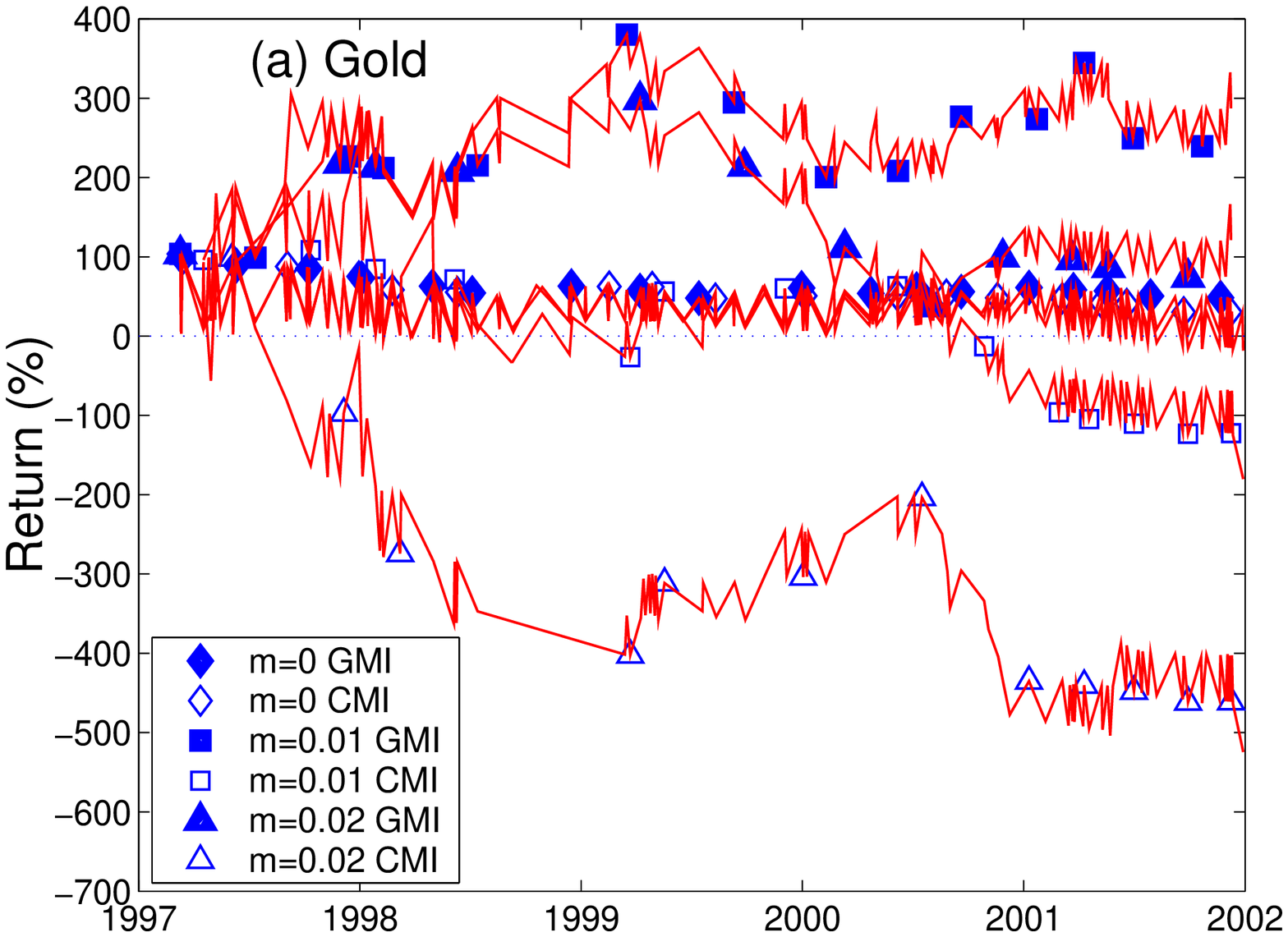} \hfill
\includegraphics[width=.48\textwidth]{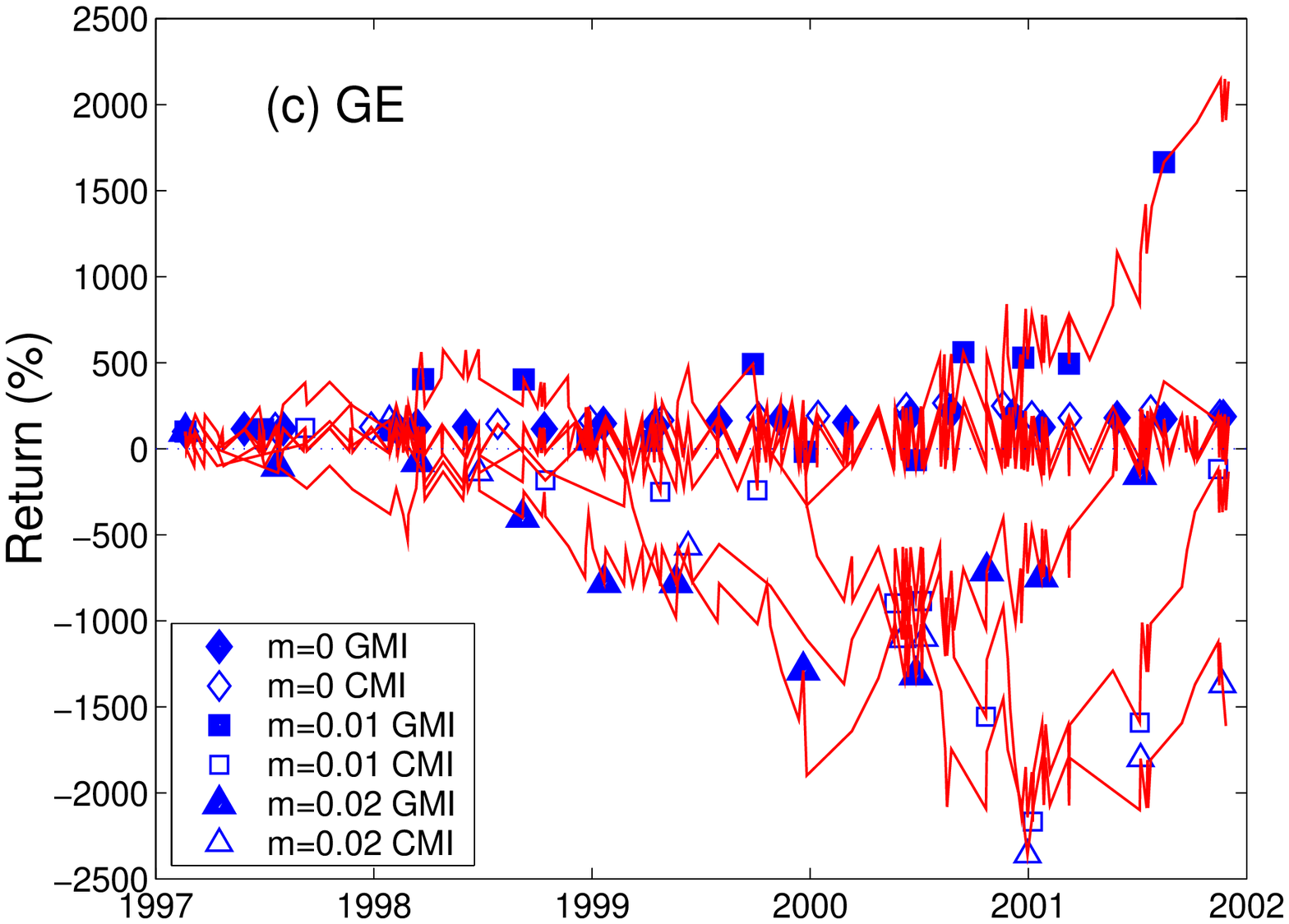} \vfill
\includegraphics[width=.48\textwidth]{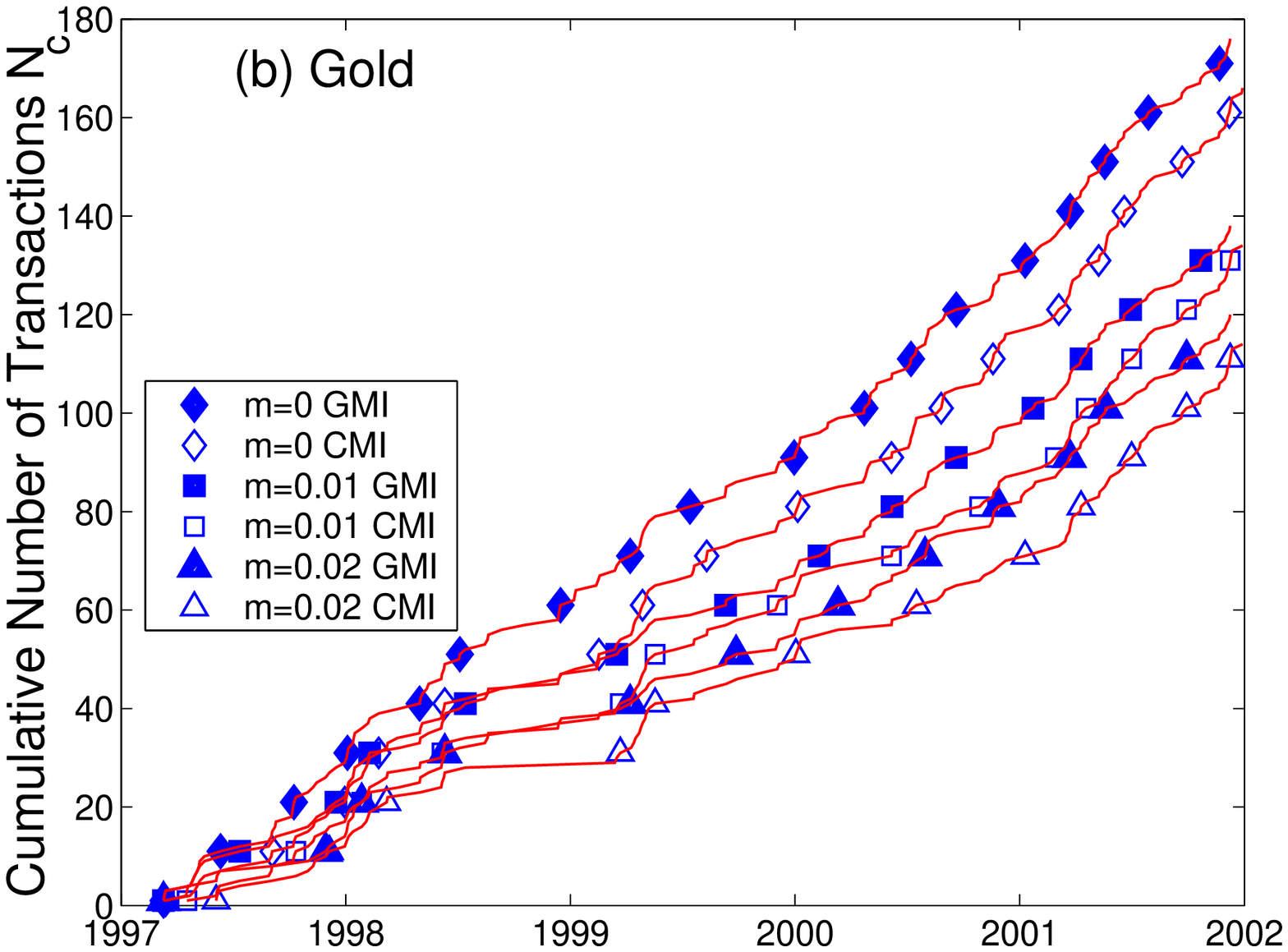} \hfill
\includegraphics[width=.48\textwidth]{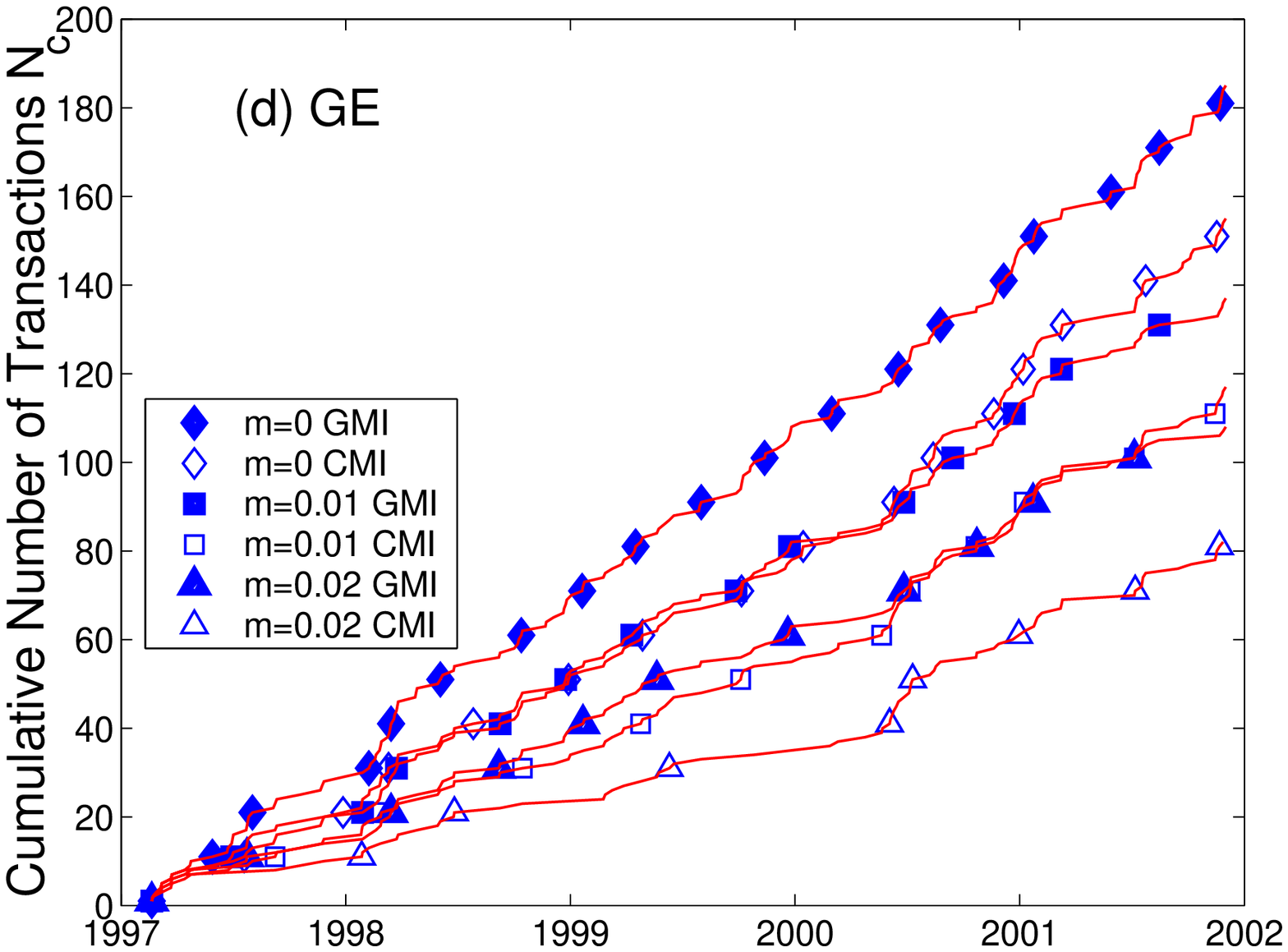} \caption{Comparison of
returns (Eq.(\ref{return})) in per cent between the generalized momentum
indicator GMI-strategy and classical momentum indicator CMI-strategy for
different values of $m$-criterion for (a) Gold and (c) GE.  Comparison of the
cumulative number of recommended transactions $N_c$ performed when GMI-strategy
and CMI-strategy are implemented for different values of $m$-criterion for (a)
Gold and (c) GE } \label{eps4} \end{figure}

Following basic logics  we {\it buy} at the minima of the GMI and {\it sell} at
the maxima of GMI. We assume that we owe one share of a given stock at time
$t_0$, which is the first {\it sell} signal of a stock $y(t_0)$. The accumulated
return that one can obtain during a certain period of time can be defined as

\begin{equation} Z(t) = \sum_{i=1}^{N_c} I(i)y(i)/y(t_0) \label{return}
\end{equation} where $N_c$ is the total number of transactions suggested by the
strategy during the period of interest; $I(i)= - 1$ at the minima of
$R^{\Sigma}_{\tau}(t)$ or $\widetilde R^{\Sigma}_{\tau}(t)$ is producing a {\it
buy} signal to investors, and $I(i)= + 1$ at the maxima of $R^{\Sigma}_{\tau}(t)$
or $\widetilde R^{\Sigma}_{\tau}(t)$ is communicating a {\it sell} signal.
Results of investments are plotted in Fig. 4 (a,c) for Gold and GE stocks
following GMI- (full diamond) and CMI-strategy (open diamond).

\begin{figure}[ht] \centering
\includegraphics[width=.48\textwidth]{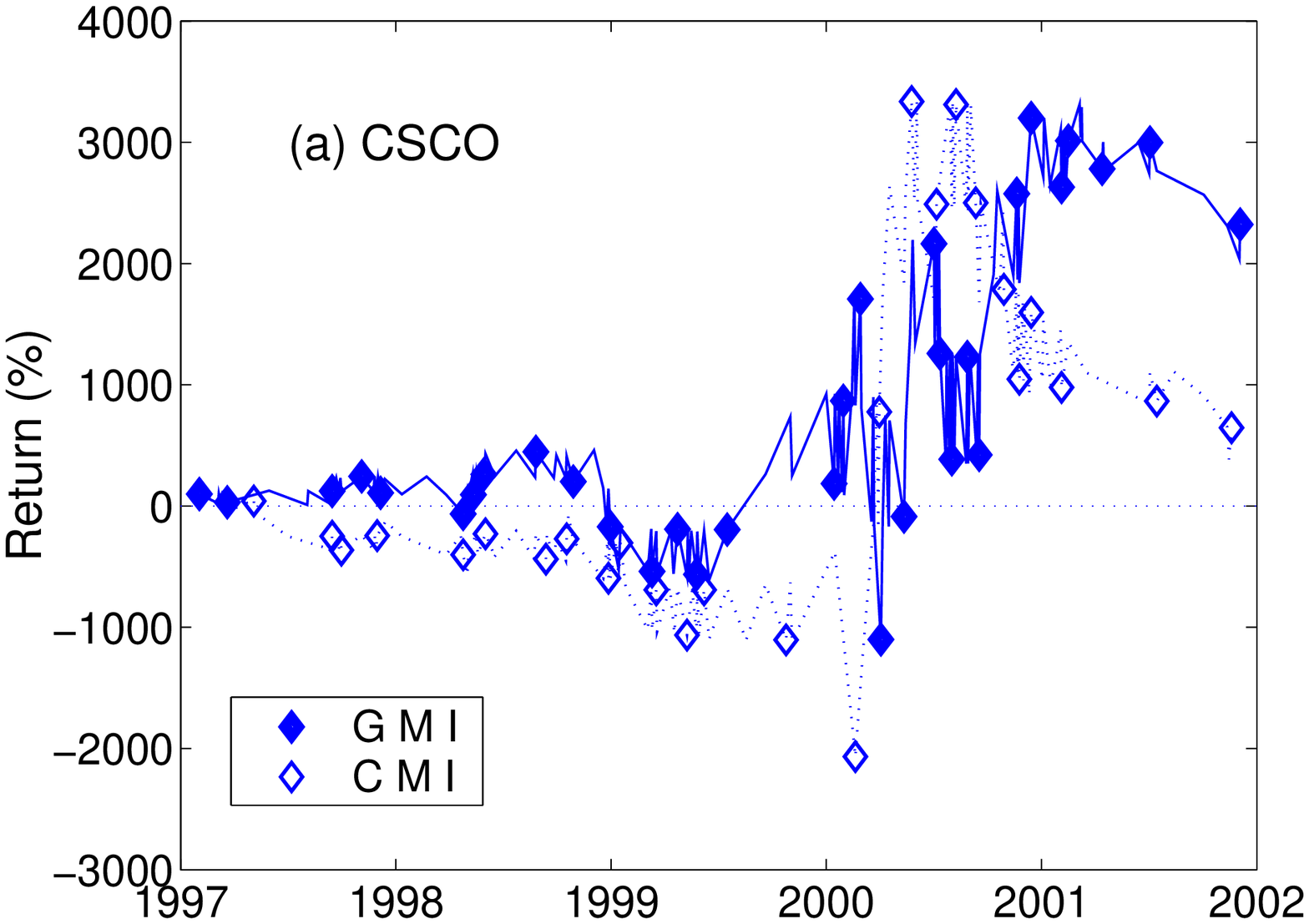} \hfill
\includegraphics[width=.47\textwidth]{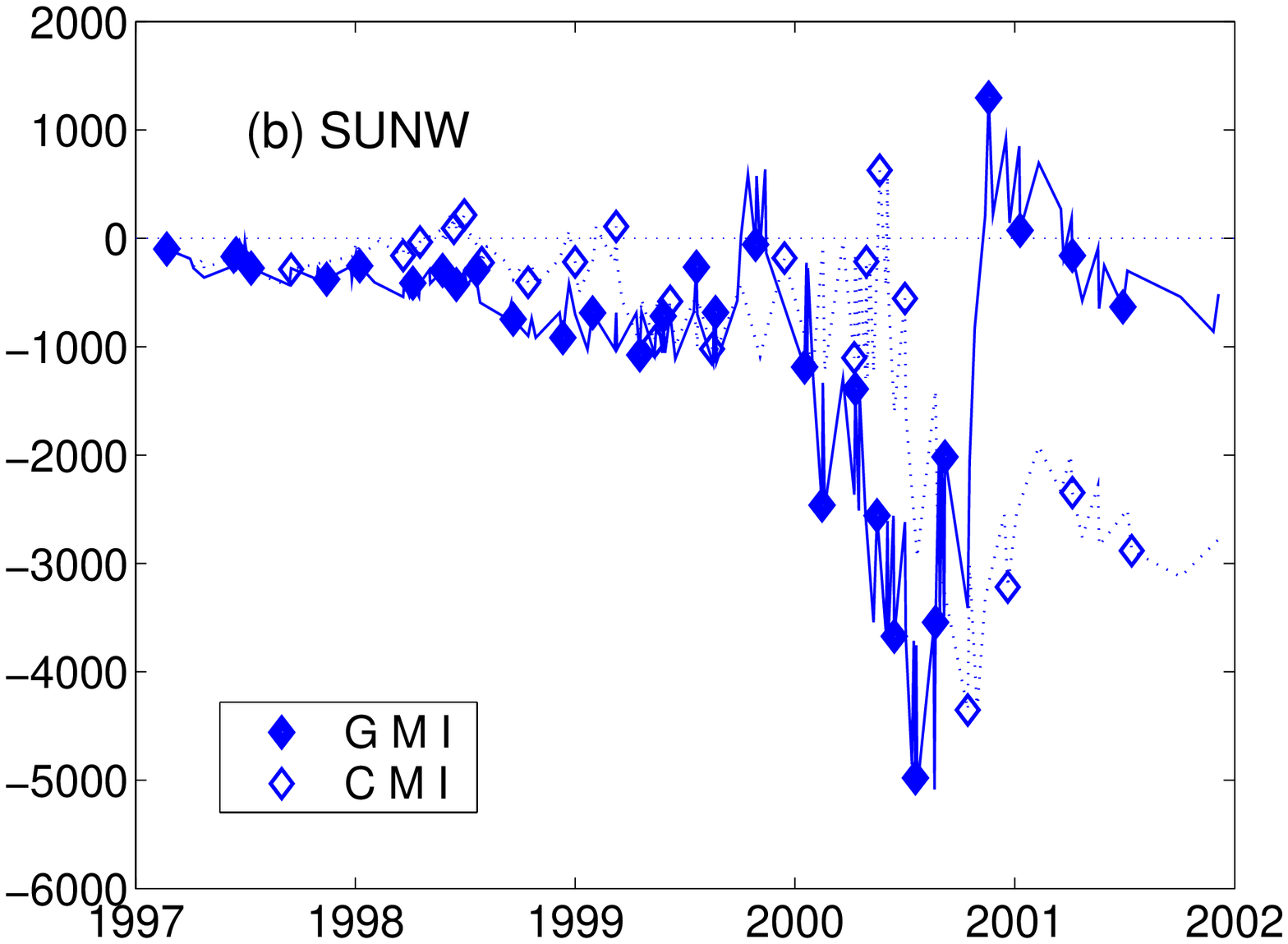} \vfill
\includegraphics[width=.48\textwidth]{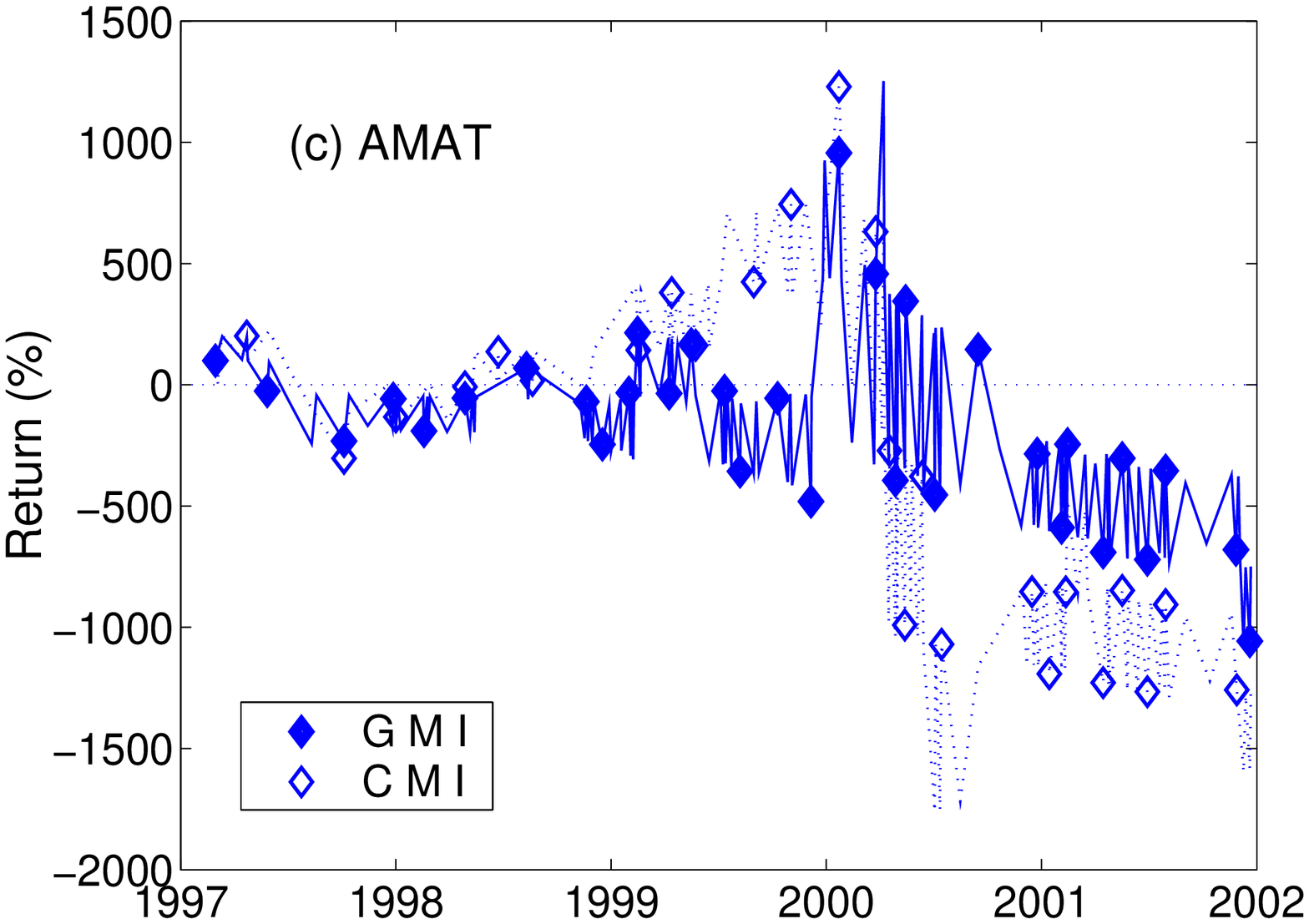} \hfill
\includegraphics[width=.47\textwidth]{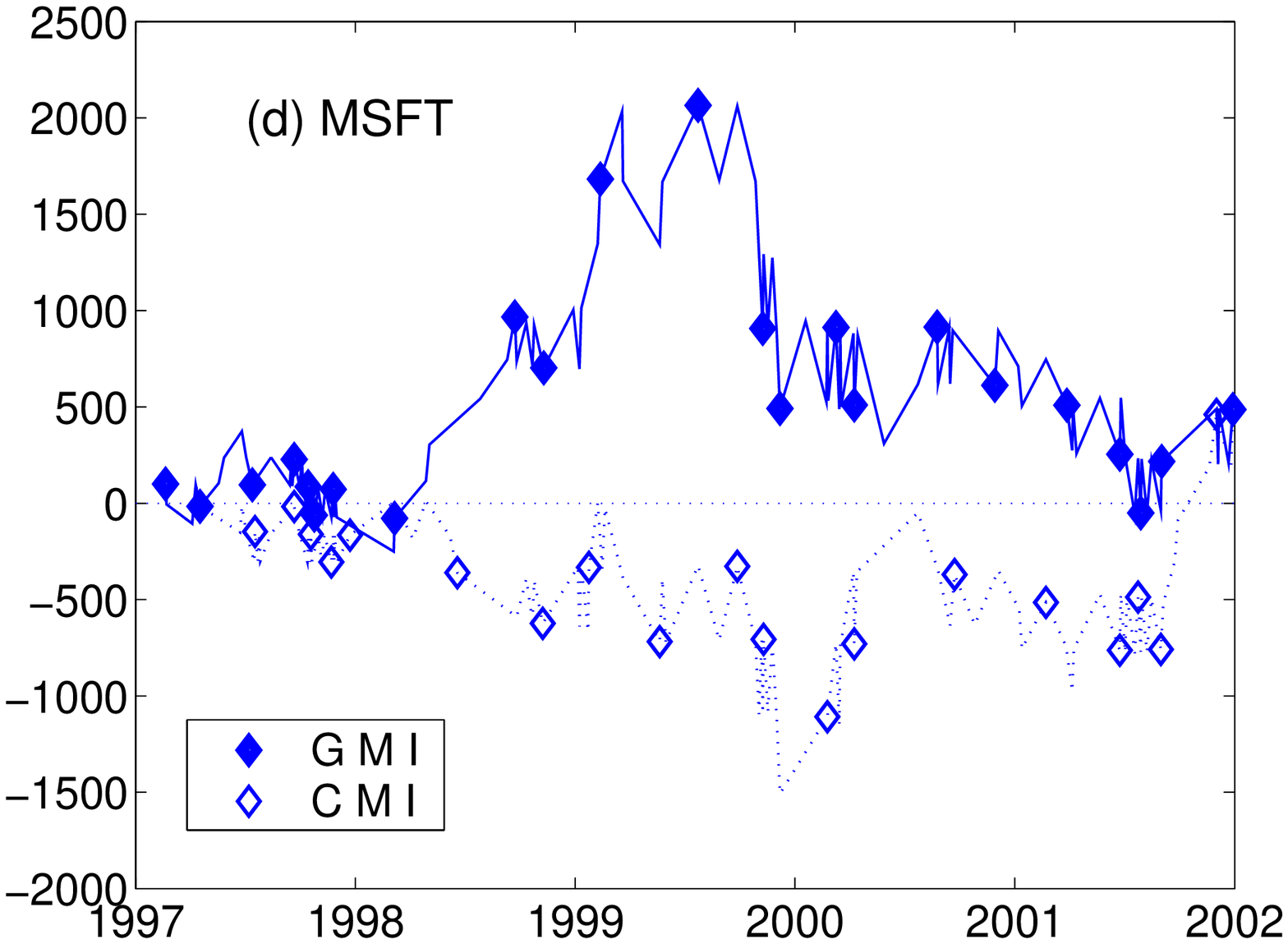}
\caption{Retuns obtained following (GMI)-strategy (full diamonds and solid line)
and CMI-strategy (open diamonds and dotted line) 
for $\tau$=21~days and restriction policy parameter $m=$0.01
for the period Jan. 1, 1997 to Dec. 31,
2001 for stocks traded on NASDAQ -- (a) CSCO, (b) SUNW, (c) AMAT, (d) MSFT}
\label{eps5} \end{figure}

\begin{figure}[ht] \centering
\includegraphics[width=.48\textwidth]{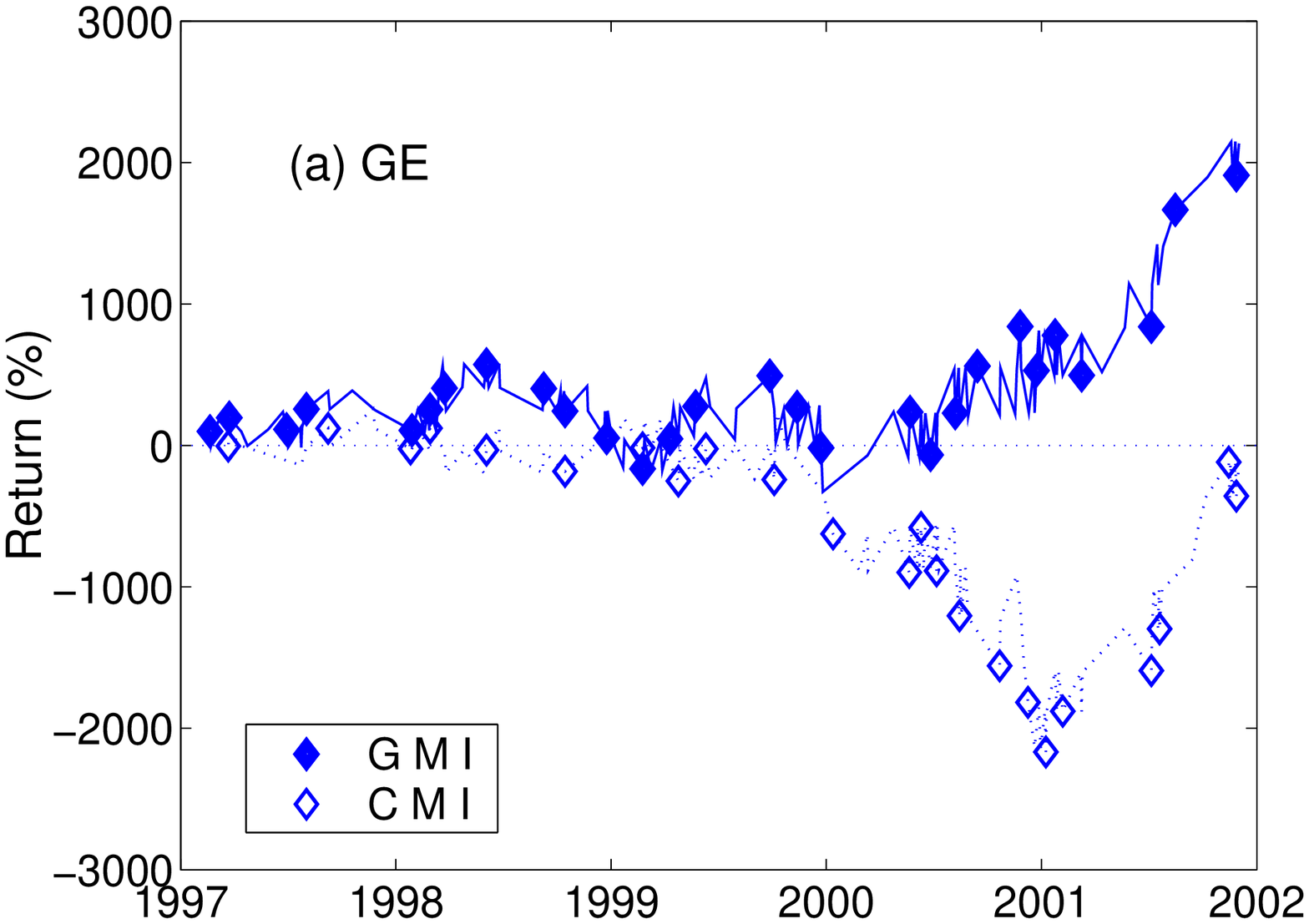} \hfill
\includegraphics[width=.46\textwidth]{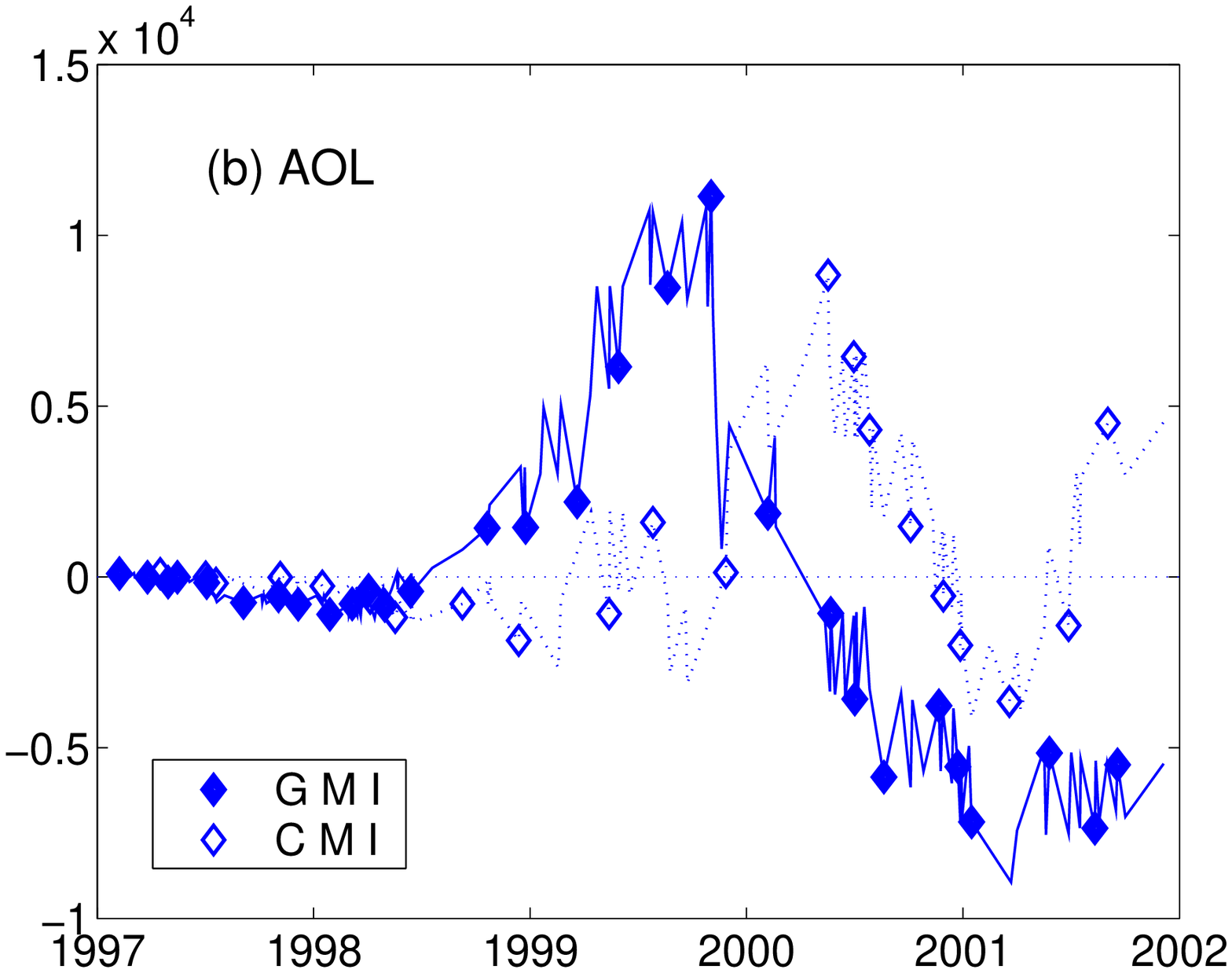} \vfill
\includegraphics[width=.48\textwidth]{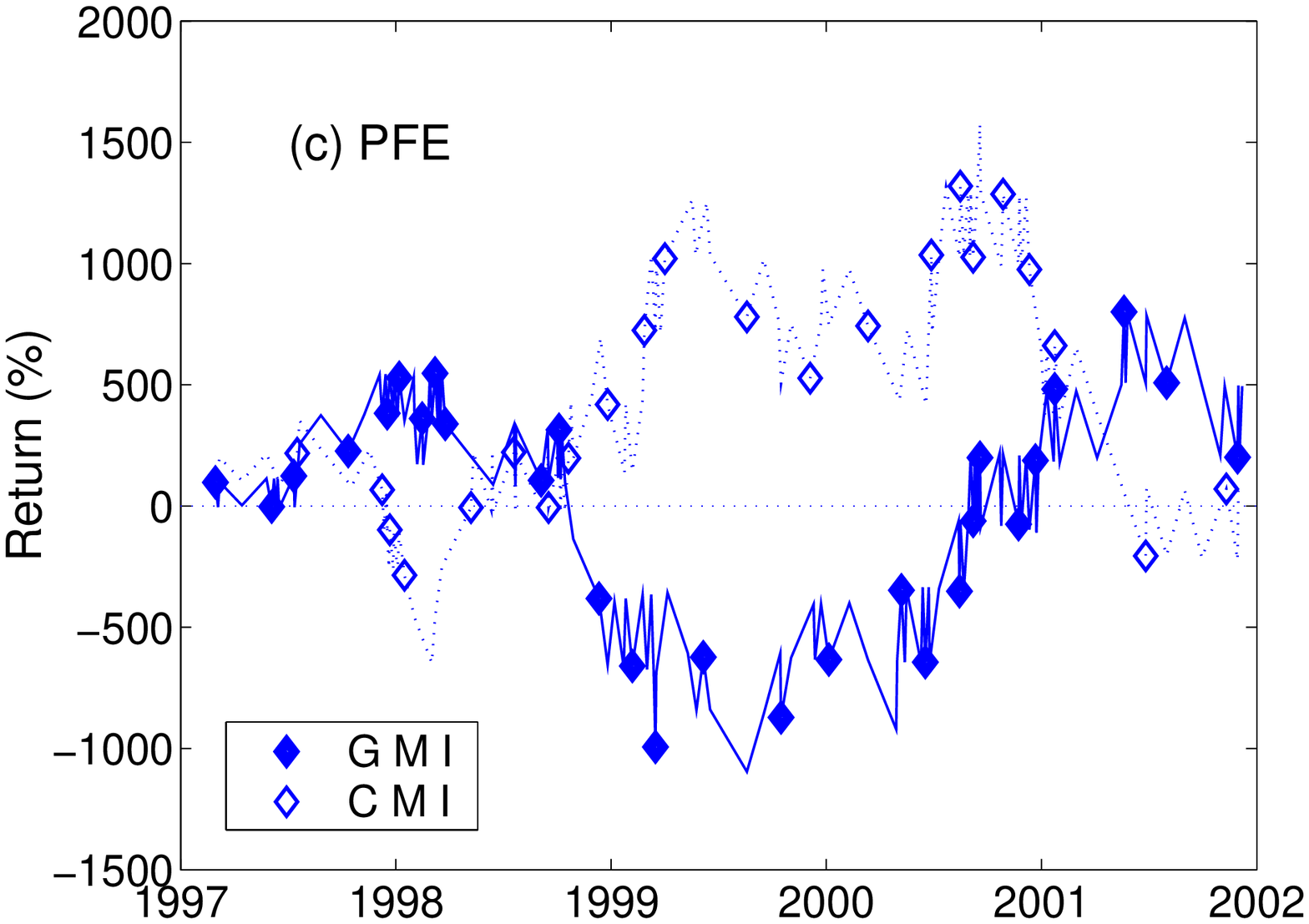} \hfill
\includegraphics[width=.46\textwidth]{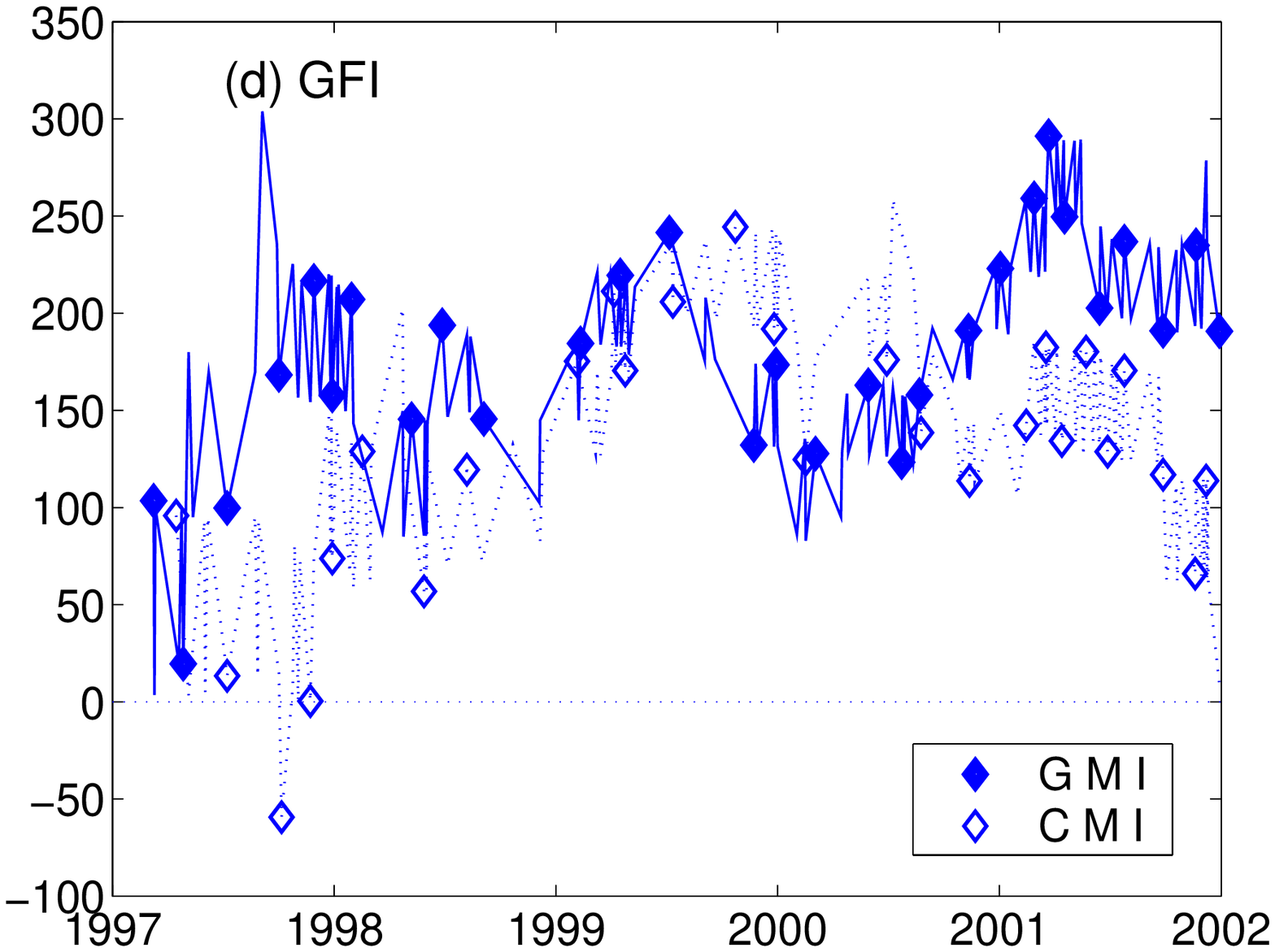} 
\caption{Retuns obtained following (GMI)-strategy (full diamonds and solid line)
and CMI-strategy (open diamonds and dotted line)
for $\tau$=21~days and restriction policy parameter $m=$0.01
for the period Jan. 1, 1997 to Dec. 31,
2001 for stocks traded on NYSE -- (a) GE, (b) AOL, (c) PFE, (d) GFI}
\label{eps6} \end{figure}

In order to take advantage of the knowledge of the dynamics of the market
represented by the volume of transactions and included in the GMI we introduce a
criterion that will prevent from investment activity (and thus monitor the risk) when the change of the
momentum indicator at the local extremum is smaller than a certain percentage of
the momentum indicator value at the time just before the extremum. Thus we define
a $m$-criterion for measuring the relative depth of the local extrema as

\begin{figure}[ht] \centering
\includegraphics[width=.48\textwidth]{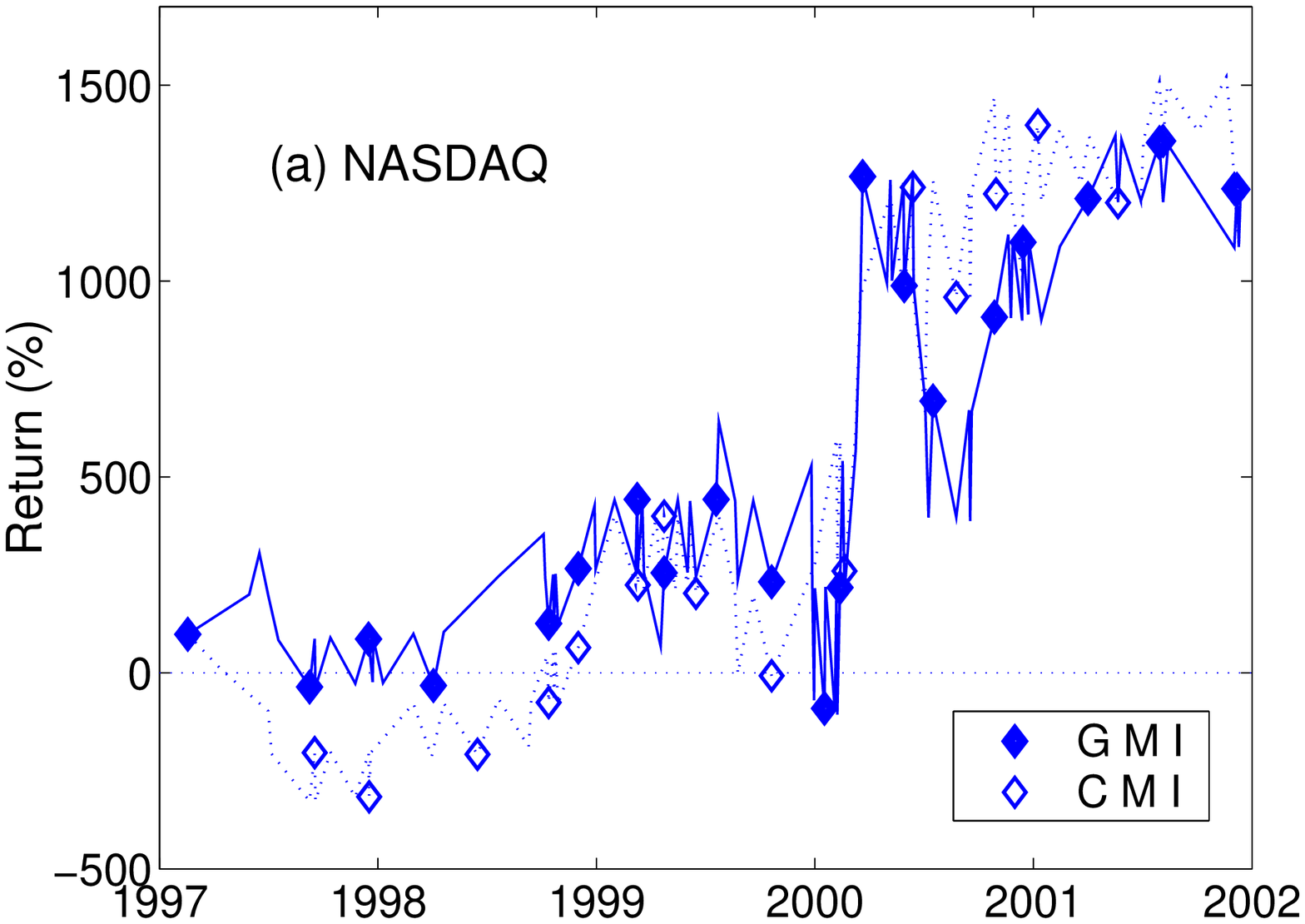} \hfill
\includegraphics[width=.47\textwidth]{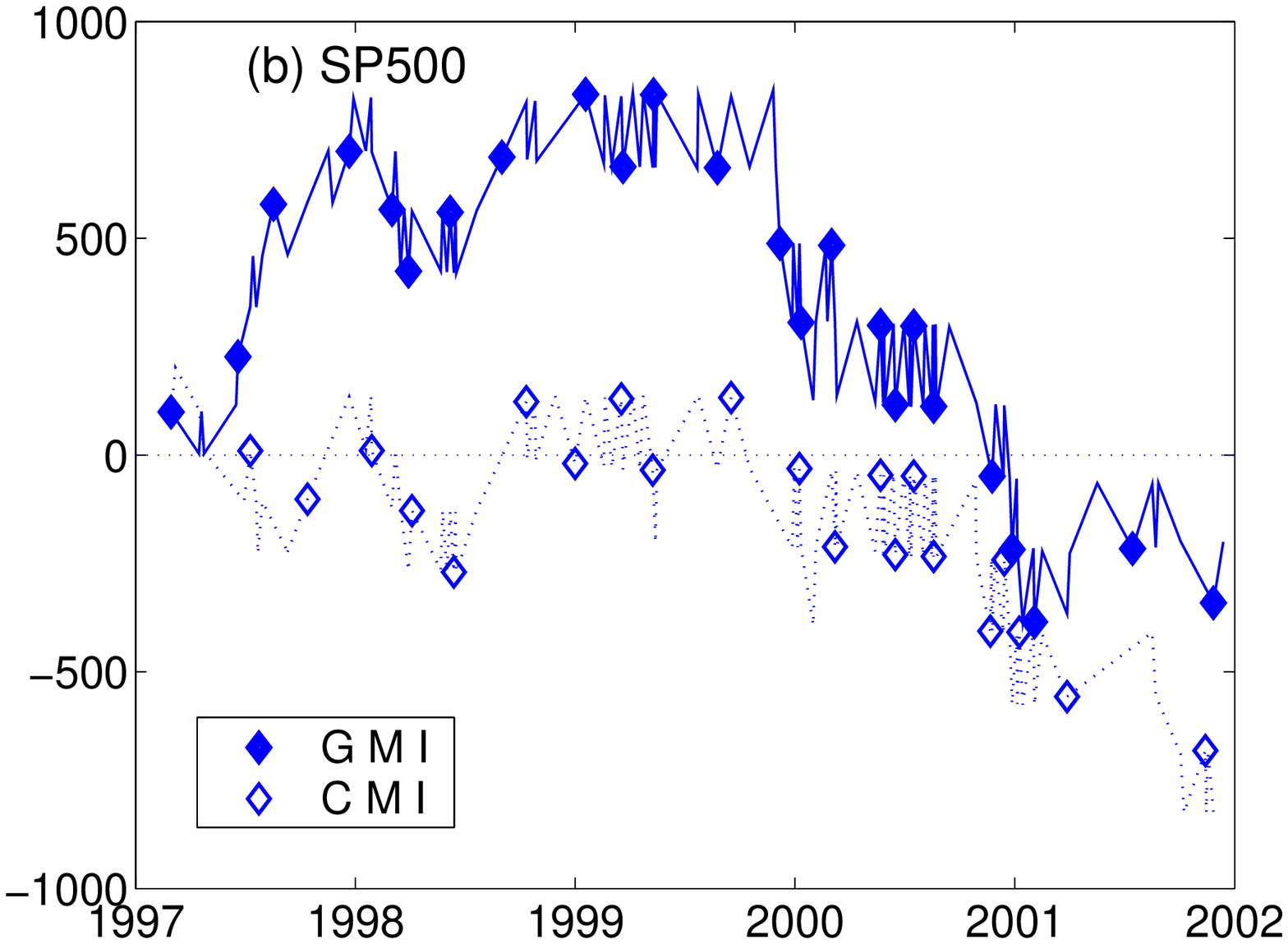} \vfill
\includegraphics[width=.48\textwidth]{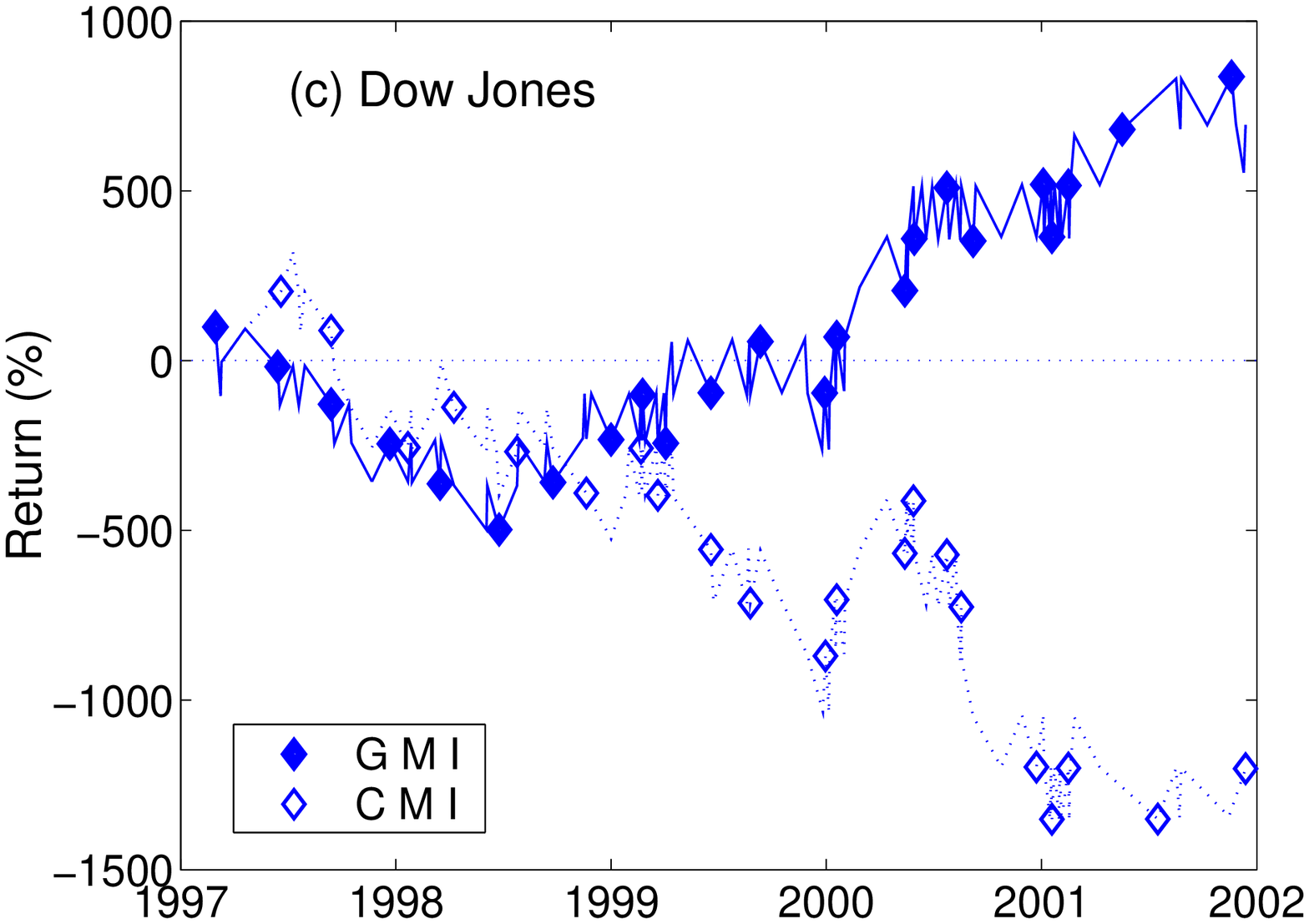} \hfill
\includegraphics[width=.47\textwidth]{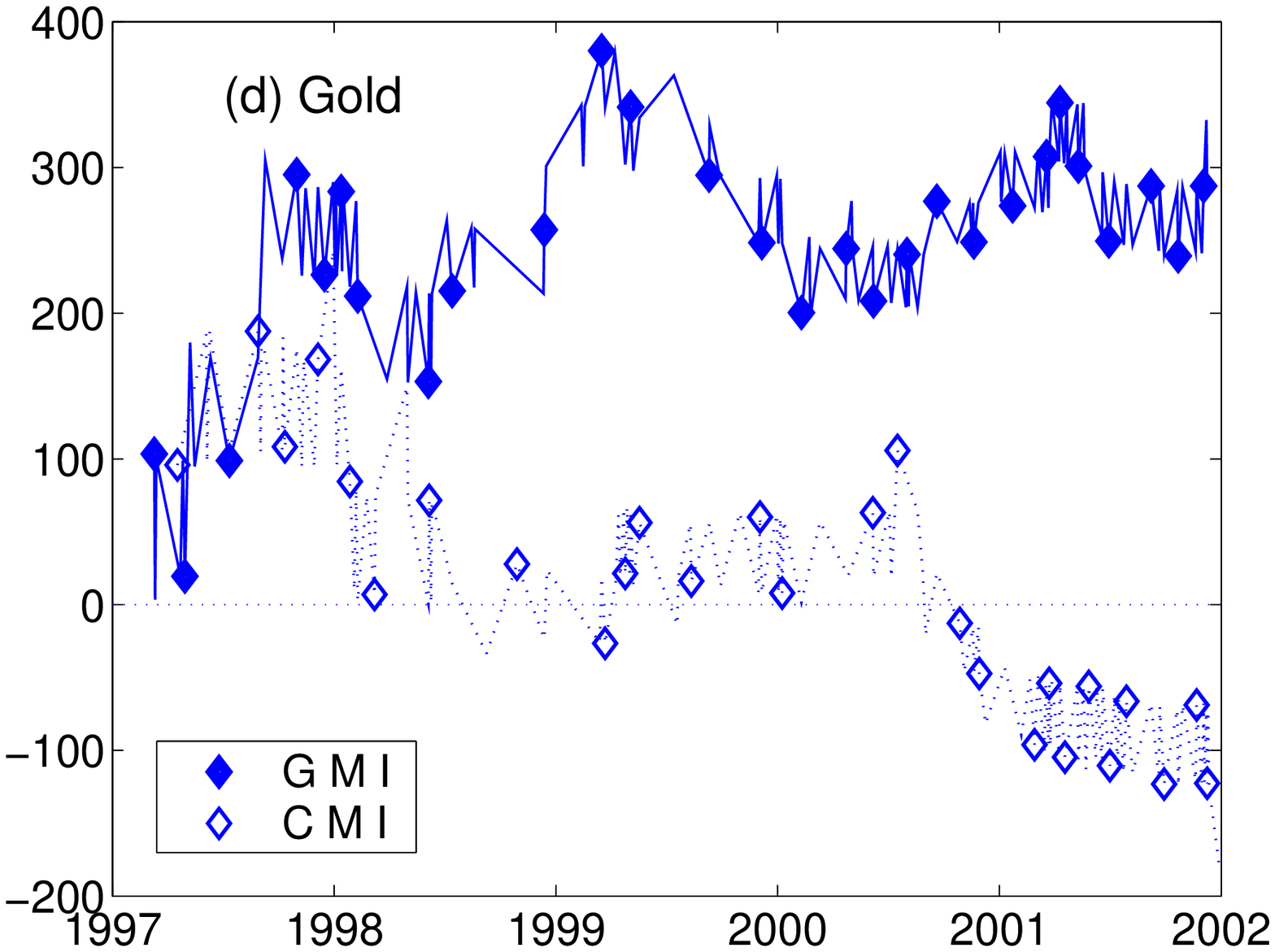} 
\caption{Retuns obtained following (GMI)-strategy (full diamonds and solid line)
and CMI-strategy (open diamonds and dotted line)
for $\tau$=21~days and restriction policy parameter $m=$0.01
for the period Jan. 1, 1997 to Dec. 31,
2001 for  three major
financial indices -- (a) NASDAQ, (b) S\&P500, (c) DJIA, the price of Gold (d)}
\label{eps7} \end{figure}

\begin{equation}  m = \left|\frac{R^{\Sigma}_{\tau}(t_{max/min}+1) -
R^{\Sigma}_{\tau}(t_{max/min})}{R^{\Sigma}_{\tau}(t_{max/min})}\right|. \label{m}
\end{equation} Different restrictions on the investment activity expressed in the
value of the $m$-criterion lead to different returns as it is illustrated in Fig.
4 (a,c) for Gold and GE, respectively. This also results in a different number of
transactions suggested by the strategies as shown in Fig. 4 (b,d) for for Gold
and GE, respectively. It is interesting to notice that for approximately the
same number of transactions, and different $m$-values the GMI-strategy leads to
higher returns. Note that an investment strategy should not be excessively
$defensive$, e.g. the returns for GE and Gold (marked with full/open triangles
(GMI/CMI)) for a relatively high (defensive) $m$=0.02 are low as compared to the
returns for $m$=0.01, marked as full/open squares (GMI/CMI). To illustrate the
performance of the GMI-strategy  with respect to the CMI-one for the stocks and
indices of interest (data in Fig. 1) we chose the $m$=0.01 investment strategy
and plot all the results in Figs. 5, 6 and 7.

\section{Conclusions}

We present an investment strategy based on a generalized momentum indicator (GMI)
that takes into account complete information from the two faceted market - the
share price and the volume of transactions. In addition, our GMI-strategy
involves restriction policies defined by the value of a hereabove introduced
$m$-criterion, describing a more or less risky investor. By doing so we obtain
profits with the GMI that are higher than those obtained with the (classical
momentum indicator) CMI-strategy with the same restriction policies applied.

\vspace*{0.2cm}

{\noindent \bf Acknowledgements} 
We are very grateful to the
organizers of the Symposium for their invitation and to the
Symposium sponsors for financial support.


\clearpage \addcontentsline{toc}{section}{Index} \flushbottom \printindex

\end{document}